\documentclass[11pt,a4paper,UKenglish,texlive=2016]{article}
\pdfoutput=1
\usepackage{jinstpub}

\usepackage[biblatex=false]{atlaspackage}

\graphicspath{{../logos/}{figures/}}

\hypersetup{pdftitle={ATLAS draft},pdfauthor={The ATLAS Collaboration}}

\usepackage{multirow}

\title{Beam test measurements of Low Gain Avalanche Detector single pads and arrays for the ATLAS High Granularity Timing Detector}

\author[a]{C.~Allaire,}
\author[b]{J.~Benitez,}
\author[c]{M.~Bomben,}
\author[c]{G.~Calderini,}
\author[d]{M.~Carulla,}
\author[e]{E.~Cavallaro,}
\author[a]{A.~Falou,}
\author[d]{D.~Flores,}
\author[f]{P.~Freeman,}
\author[f]{Z.~Galloway,}
\author[e,f]{E.L.~Gkougkousis,}
\author[f]{H.~Grabas,}
\author[e]{S.~Grinstein,}
\author[f]{B.~Gruey,}
\author[g]{S.~Guindon,}
\author[g]{A.M.~Henriques~Correia,}
\author[d]{S.~Hidalgo,}
\author[h]{A.~Kastanas,}
\author[f]{C.~Labitan,}
\author[c]{D.~Lacour,}
\author[e]{J.~Lange,}
\author[i]{F.~Lanni,}
\author[g]{B.~Lenzi,}
\author[f]{Z.~Luce,}
\author[a]{N.~Makovec,}
\author[c]{G.~Marchiori,}
\author[j,1]{L.~Masetti,%
\note{Corresponding author.}}
\author[d]{A.~Merlos,}
\author[f]{F.~McKinney-Martinez,}
\author[c]{I.~Nikolic-Audit,}
\author[d]{G.~Pellegrini,}
\author[g]{R.~Polifka,}
\author[d]{D.~Quirion,}
\author[g]{A.~Rummler,}
\author[f]{H.F-W.~Sadrozinski,}
\author[f]{A.~Seiden,}
\author[a]{L.~Serin,}
\author[a]{S.~Simion,}
\author[f]{E.~Spencer,}
\author[c]{S.~Trincaz-Duvoid,}
\author[f]{M.~Wilder,}
\author[f]{A.~Zatserklyaniy,}
\author[a]{D.~Zerwas}
\author[f]{and Y.~Zhao.}

\affiliation[a]{LAL, Univ. Paris-Sud, CNRS/IN2P3, Universit{\'e} Paris-Saclay, Orsay, France}
\affiliation[b]{University of Iowa, Iowa City IA, United States of America}
\affiliation[c]{Laboratoire de Physique Nucl{\'e}aire et de Hautes Energies, UPMC and Universit{\'e} Paris-Diderot and CNRS/IN2P3, Paris, France}
\affiliation[d]{Centro Nacional de Microelectronica (CNM-IMB-CSIC), Campus UAB, 08193 Bellaterra (Barcelona), Spain}
\affiliation[e]{Institut de F{\'\i}sica d'Altes Energies (IFAE), The Barcelona Institute of Science and Technology, Barcelona, Spain}
\affiliation[f]{Santa Cruz Institute for Particle Physics, University of California Santa Cruz, Santa Cruz CA, United States of America}
\affiliation[g]{CERN, Geneva, Switzerland}
\affiliation[h]{Physics Department, Royal Institute of Technology, Stockholm, Sweden}
\affiliation[i]{Physics Department, Brookhaven National Laboratory, Upton NY, United States of America}
\affiliation[j]{Institut f{\"u}r Physik, Universit{\"a}t Mainz, Mainz, Germany}

\emailAdd{Lucia.Masetti@cern.ch}

\abstract{%
For the high luminosity upgrade of the LHC at CERN, ATLAS is considering the addition of a High Granularity Timing Detector (HGTD) in front of the end cap and forward calorimeters at $|z|= 3.5$ m and covering the region $2.4 <\eta< 4$ to help reducing the effect of pile-up. The chosen sensors are arrays of  50~\SI{}{\micro\meter} thin Low Gain Avalanche Detectors (LGAD). This paper presents results on single LGAD sensors with a surface area of 1.3$\times$1.3\,mm$^2$ and arrays with 2$\times$2 pads with a surface area of 2$\times$2\,mm$^2$ or 3$\times$3\,mm$^2$ each and different implant doses of the p$^+$ multiplication layer. They are obtained from data collected during a beam test campaign in autumn 2016 with a pion beam of 120\,GeV energy at the CERN SPS. In addition to several quantities measured inclusively for each pad, the gain, efficiency and time resolution have been estimated as a function of the position of the incident particle inside the pad by using a beam telescope with a position resolution of few $\mu$m. Different methods to measure the time resolution are compared, yielding consistent results. The sensors with a surface area of 1.3$\times$1.3\,mm$^2$ have a time resolution of about 40\,ps for a gain of 20 and of about 27\,ps for a gain of 50 and fulfil the HGTD requirements. Larger sensors have, as expected, a degraded time resolution.  All sensors show very good efficiency and time resolution uniformity.%
}


\arxivnumber{1804.00622}

\begin{document}

\maketitle

\section{Introduction}
\label{sec:intro}

The correct assignment of the particles originating from the hard-scattering process and the suppression of detector signals produced by a nominal average of 200 additional low-energy $pp$ collisions (pile-up) are among the most difficult challenges at the high luminosity upgrade of the Large Hadron Collider (HL-LHC) at CERN.
A High Granularity Timing Detector (HGTD) in the end-cap/forward region of the ATLAS detector \cite{PERF-2007-01}, covering $2.4<|\eta|<4.0$, adds capabilities with respect to the foreseen new inner tracker to mitigate these effects on physics final states containing forward jets. Due to the high radiation levels expected in this region for an integrated luminosity of $\mathcal{L}=4000$ fb$^{-1}$, the detector sensors and front-end electronics must sustain a 1 MeV neutron equivalent fluence of up to 3.7$\times$10$^{15}$ neutrons/cm$^{2}$ ~and 4.5\,MGy total ionising dose (at R=120\,mm, including a safety factor of 1.5 and assuming one replacement of the inner part after half of the lifetime), while providing the challenging time resolution requirements of approximately 30 ps per minimum ionising particle (MIP). The sensor choice for the HGTD, given the need for accurate time measurement, are Low Gain Avalanche Detector (LGAD) pads with a thickness of about 50~\SI{}{\micro\meter} and a pad area of 1.3$\times$1.3\,mm$^2$.

In autumn 2016 single pads and arrays of LGADs mounted on custom electronic boards providing amplification were tested with a high-energy pion beam at the H6B line at the CERN SPS. The obtained results are reported in this paper. Previous results for single sensors before and after irradiation can be found in Refs.~\cite{bib:UFSD300umTB,bib:UFSD50umTBNicolo,bib:AFPLGAD,Galloway:2017gfx,bib:thinLGADradiationHardness,Zhao:2018qkg}. In this paper, the previous results have been confirmed and extended to results of LGAD arrays, including uniformity scans of the pad surface.

The time resolution of the detector is given by the quadratic sum of the dispersion due to non-uniform energy deposition along the sensor causing fluctuations in the Landau distribution ($\sigma_{\mathrm{Landau}}$) and of the electronic noise ($\sigma_{\mathrm{elec}}$), which is dominated by two effects: jitter ($\sigma_\textrm{jitter}$) and time walk ($\sigma_\textrm{time\,walk}$). While the Landau term can be reduced by using thin sensors, both terms of the electronic noise depend inversely on the signal slope $dV$/$dt$:
\begin{equation}
        \sigma_\textrm{jitter}=\frac{N}{(dV/dt)}  \simeq \frac{t_\textrm{rise}}{(S/N)}, \hspace{2cm}
        \sigma _\textrm{time walk} =  \left[\frac{V_\textrm{th}}{\frac{S}{t_\textrm{rise }}}\right]_\textrm{RMS} \propto \left[\frac{N}{\frac{dV}{dt}}\right]_\textrm{RMS}
\label{equa:TWjitter}
\end{equation}
In Equation \ref{equa:TWjitter}, $N$ is the electronic noise, $t_\textrm{rise}$ the rise time for the signal, $S$ the signal amplitude and $V_\textrm{th}$ the voltage used as threshold to determine the time of arrival.
An additional term in the time resolution of the final detector depends on the size of the time-to-digital converter (TDC) bin, but this is not considered for the present results, since instead of a TDC an oscilloscope with a very high sampling rate was used. Similarly, the precision of the clock distribution is relevant for the time resolution of the final detector, but not for the results of this paper.

This paper is organised as follows: In Section \ref{sec:senselec} the devices under test as well as the read-out electronics are described, while the beam test setup and the data acquisition system are presented in Section \ref{sec:setup}. The data reconstruction and analysis methods are detailed in Section \ref{sec:reco}, followed by the presentation and discussion of the results in Section \ref{sec:results}.


\section{Sensors and electronics}
\label{sec:senselec}

In this Section, the main features of the devices under test as well as results from laboratory studies are presented. A summary of the tested sensors and used read-out boards with the naming conventions used throughout this paper can be found in Table~\ref{tab:deviceTable} at the end of the Section.

\begin{figure}[hbtp]
	\centering
	\includegraphics[width=14cm]{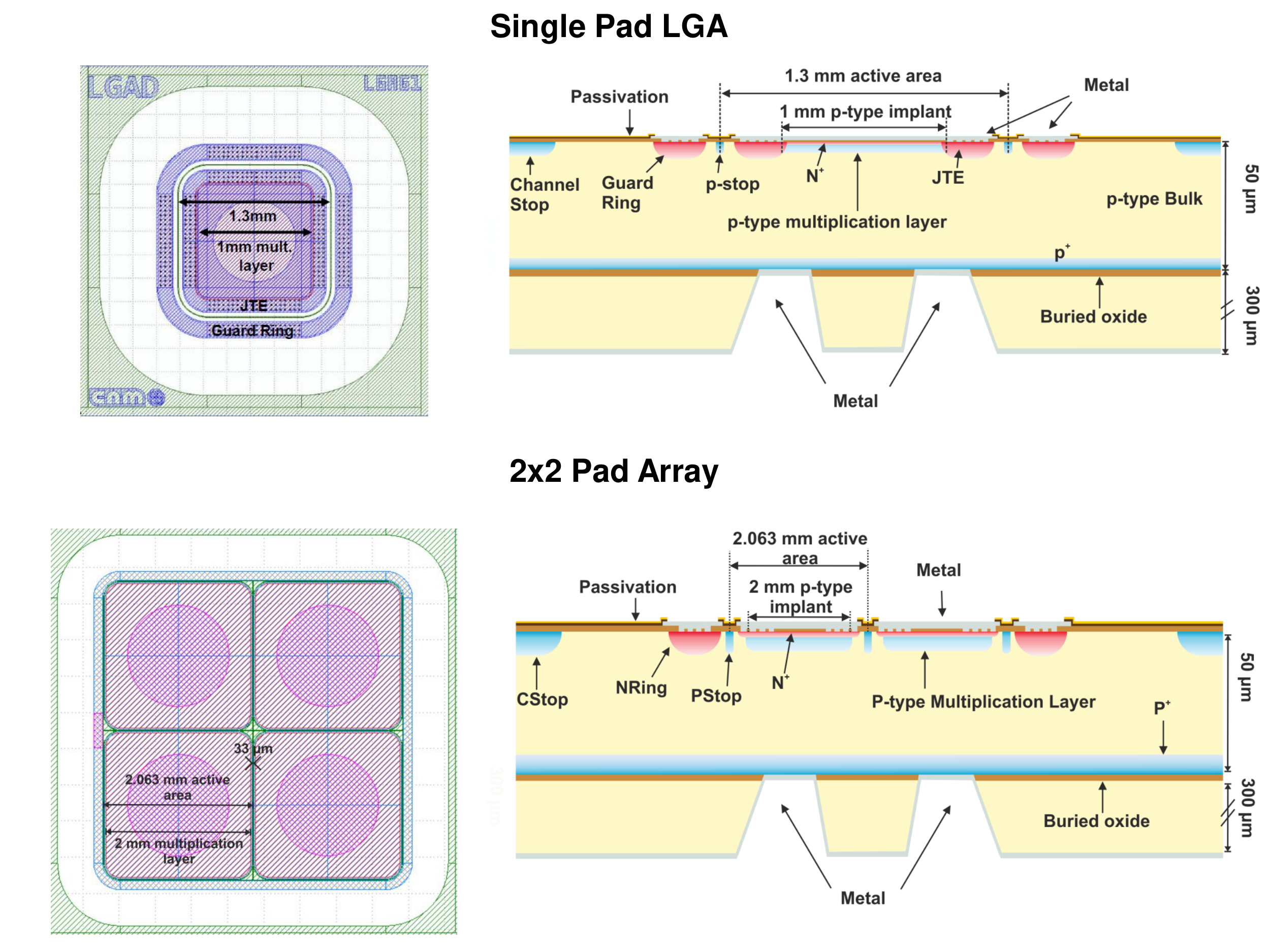}
	\caption{Sketches of the single pad (top) and 2$\times$2 array LGAD sensors. A top view is shown on the left and a side view on the right.}
	\label{fig:samples}
\end{figure}

\subsection{LGAD Sensors}
The sensors are thin pixelated n-on-p silicon sensors whose geometry has been optimized for precision time measurements. They are based on the LGAD design~\cite{bib:LGAD,bib:MarTorino} developed by the Centro Nacional de Microelectronica (CNM) Barcelona within a RD50 Common Project\footnote{\url{http: //rd50.web.cern.ch/rd50/}} (Run 9088). LGADs are based on implanting a few micrometer thick highly doped p-type layer between the high resistivity p-type bulk and the n$^+$ implant, which acts as a high-field charge multiplication layer providing moderate gain of about 5--70. The devices studied here are produced on 4" silicon-on-insulator (SOI) wafers with nominally 50~\SI{}{\micro\meter} thickness and 12\,k$\Omega$cm resistivity on a 300~\SI{}{\micro\meter} thick support wafer and 1~\SI{}{\micro\meter} buried oxide. Boron is used as dopant for the p-type multiplication layer. Due to the diffusion of the highly doped n$^+$ and p$^+$~implants at the front and back side, respectively, the active thickness is reduced to about 45~\SI{}{\micro\meter}, which is consistent with capacitance measurements. The back-side contact is done through wet-etched deep access holes through the insulator. The wafers contain a variety of pad structures, such as single-pad diodes and segmented arrays of pad diodes with various dimensions. This test uses single pads of overall active area of 1.3$\times$1.3\,mm$^2$ (called LGA) and 2$\times$2 arrays of pads with 2.063$\times$2.063\,mm$^2$ or 3.063$\times$3.063\,mm$^2$ active area each. It should be noted that the region with charge multiplication is slightly less, namely 1.0$\times$1.0\,mm$^2$ for the LGA single pads and 2.000$\times$2.000\,mm$^2$ or 3.000$\times$3.000\,mm$^2$ for each pad of the 2$\times$2 arrays. That means there is an expected region of 63~\SI{}{\micro\meter} in between the multiplication layers of adjacent pads in the arrays. Figure~\ref{fig:samples} shows a top and a side view sketch of an LGA and a 2$\times$2\,mm$^2$ array. The single pads contain a circular opening in the top metallisation for light injection tests, whereas the array pads are fully metallised.
The single pad devices are surrounded by a deep n$^+$-implant ("NRing" or Junction Termination Extension, JTE) that protects all sides of the pad from too high fields and hence early breakdown, whereas for the arrays it surrounds only the whole structure, not each pad.
Three sets of wafers were produced, identical in the mask design but with a different multiplication layer implantation dose to optimize the gain: $1.8\times10^{13}$\,cm$^{-2}$ (low), $1.9\times10^{13}$\,cm$^{-2}$ (medium) and $2.0\times10^{13}$\,cm$^{-2}$ (high).

\begin{figure}[htbp]
  \centering
  \includegraphics[width=0.5\textwidth]{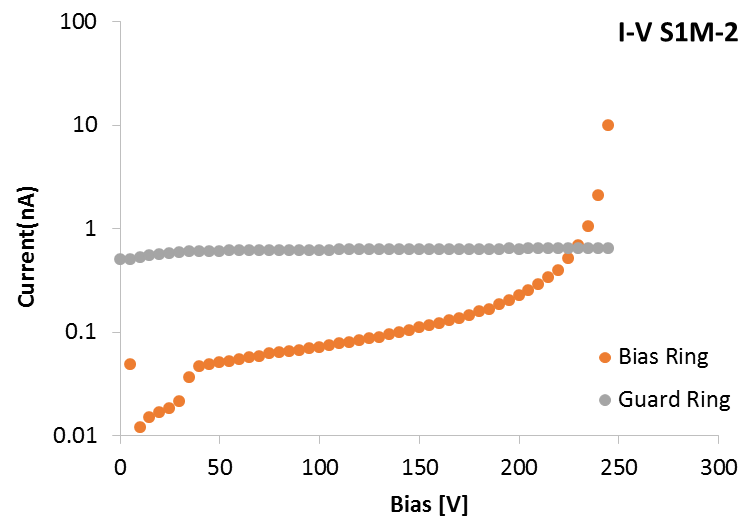}
  \caption{Current–voltage (I-V) relationship of the bias ring and guard ring, respectively, for the S1M-2 LGAD. The depletion voltage at 35\,V, the on-set of high multiplication above a bias of about 150\,V, as well as the final break down at about $V_{BD} = 250$\,V are clearly visible.}
\label{fig:IVLab}
\end{figure}

Capacitance-voltage (C-V) and current-voltage (I-V) measurements were performed in the laboratory. The C-V measurements were done with a grounded guard ring and resulted in a depletion voltage below 50\,V. The detector capacitance was measured to be C = 3.9\,pF for the LGA devices, from which the active thickness of the LGAD, $w$, was derived to be $w = 45$~\SI{}{\micro\meter} using the relationship $w= \epsilon_r \epsilon_0\cdot A/C$, where $\epsilon_0$ is the vacuum absolute permittivity, $\epsilon_r =11.7$ the silicon relative permittivity, and $A$ is the active area. The I-V measurements of the LGAD revealed a bulk leakage current of the order of 0.1\,nA with a constant guard ring current of about 1\,nA before the breakdown voltage (see Fig.~\ref{fig:IVLab}). The breakdown voltage increases with decreasing multiplication layer dose. For LGA single pad diodes about 80, 250 and 300\,V were found for high, medium and low dose, respectively. The breakdown voltage for arrays was found to be reduced (about 200\,V for medium dose) due to the absence of a JTE around each pad as explained above.

Beam tests and laboratory measurements have been performed in the past both for 300~\SI{}{\micro\meter}, 50~\SI{}{\micro\meter} and 35~\SI{}{\micro\meter} thick LGADs, mostly single-pad devices~\cite{bib:UFSD300umTB,bib:UFSD50umTBNicolo,bib:AFPLGAD,Galloway:2017gfx,bib:thinLGADradiationHardness,Zhao:2018qkg}.
In particular, results of laboratory studies on the devices of the same run used here can be found in Refs.~\cite{bib:UFSD50umTBNicolo,bib:AFPLGAD,bib:thinLGADradiationHardness}.

\subsection{Read-out boards}
\begin{figure}[htbp]
\centering
  \subfloat[]{
    \includegraphics[height=0.38\textwidth]{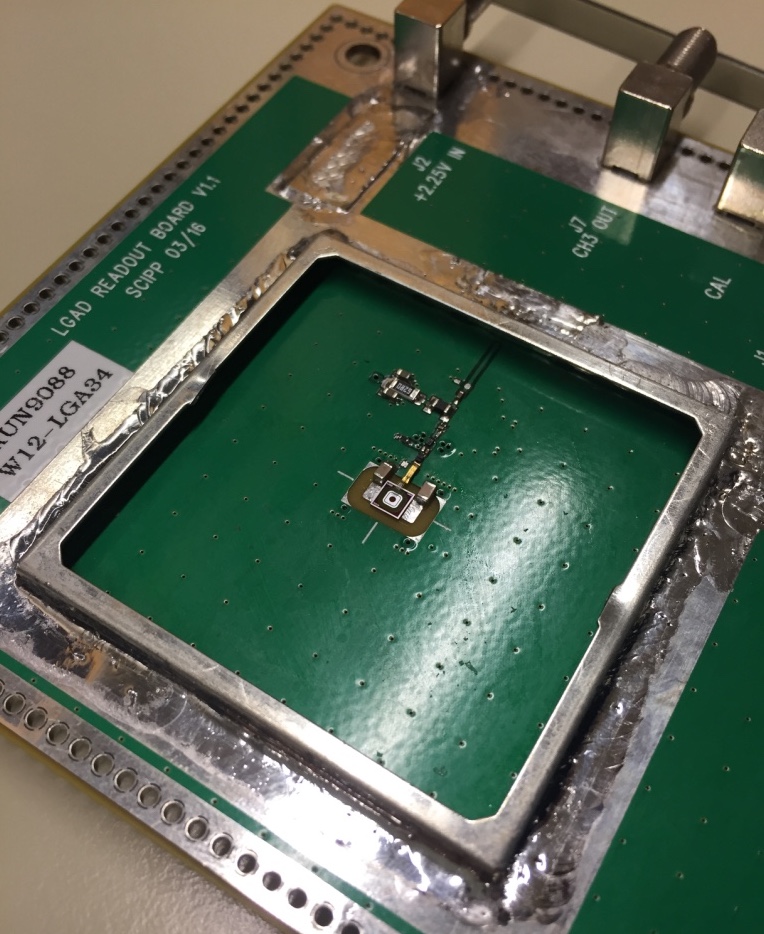}
    \label{fig:UCSC-single-board}
  }
  \hfill
  \subfloat[]{
    \includegraphics[height=0.38\textwidth]{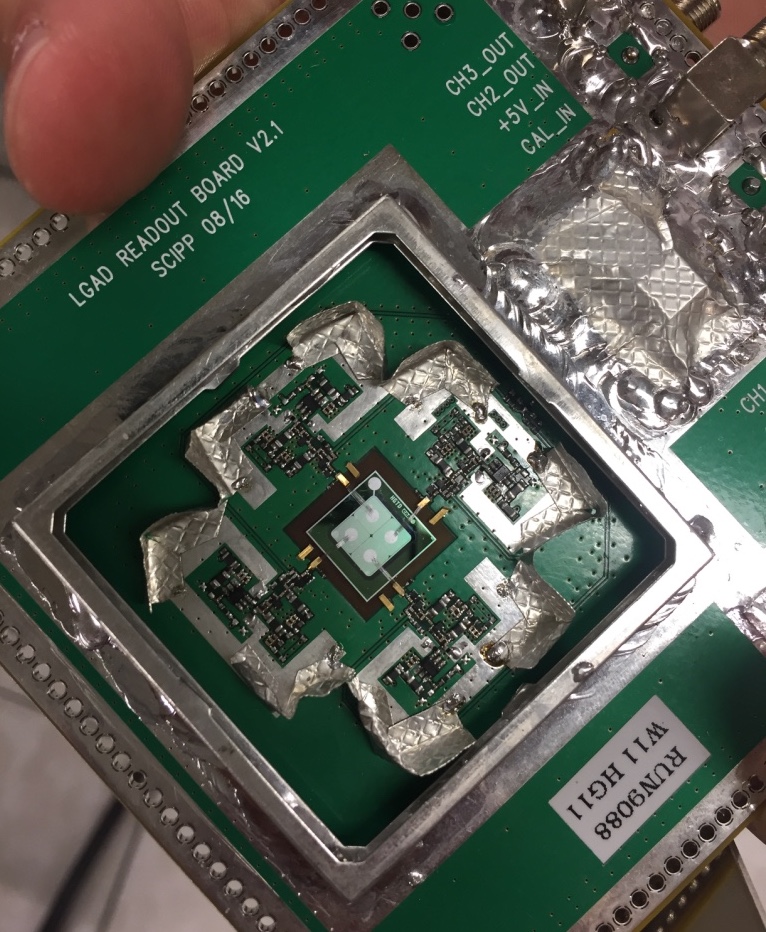}
    \label{fig:UCSC-array-board}
  }
  \hfill
  \subfloat[]{
    \includegraphics[height=0.38\textwidth]{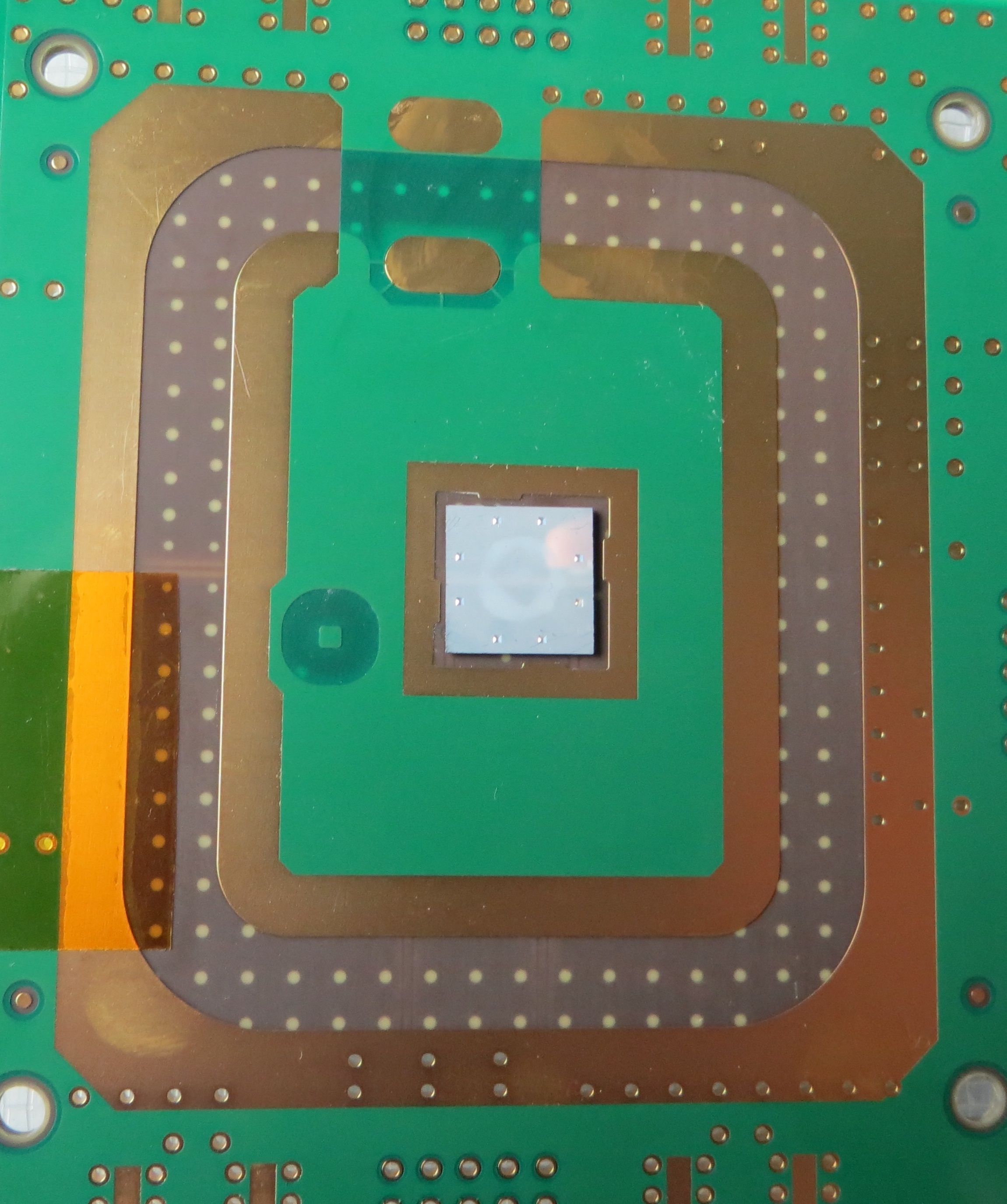}
    \label{fig:IN2P3-array-board}
  }
\caption{The three different read-out board versions used: \#1 \protect\subref{fig:UCSC-single-board}, \#2 \protect\subref{fig:UCSC-array-board}, \#3 \protect\subref{fig:IN2P3-array-board}.}
\label{fig:boards}
\end{figure}

For charge collection studies in the laboratory and the beam test measurements, the LGADs are mounted on 10$\times$10\,cm$^2$ read-out boards.
Three different versions were used: one for single pads (\#1), and two different versions of 4-channel boards for the 2$\times$2 arrays (\#2 and \#3).
They are displayed in Figure~\ref{fig:boards}.

\begin{figure}[htbp]
\centering
  \subfloat[]{
    \includegraphics[width=0.45\textwidth]{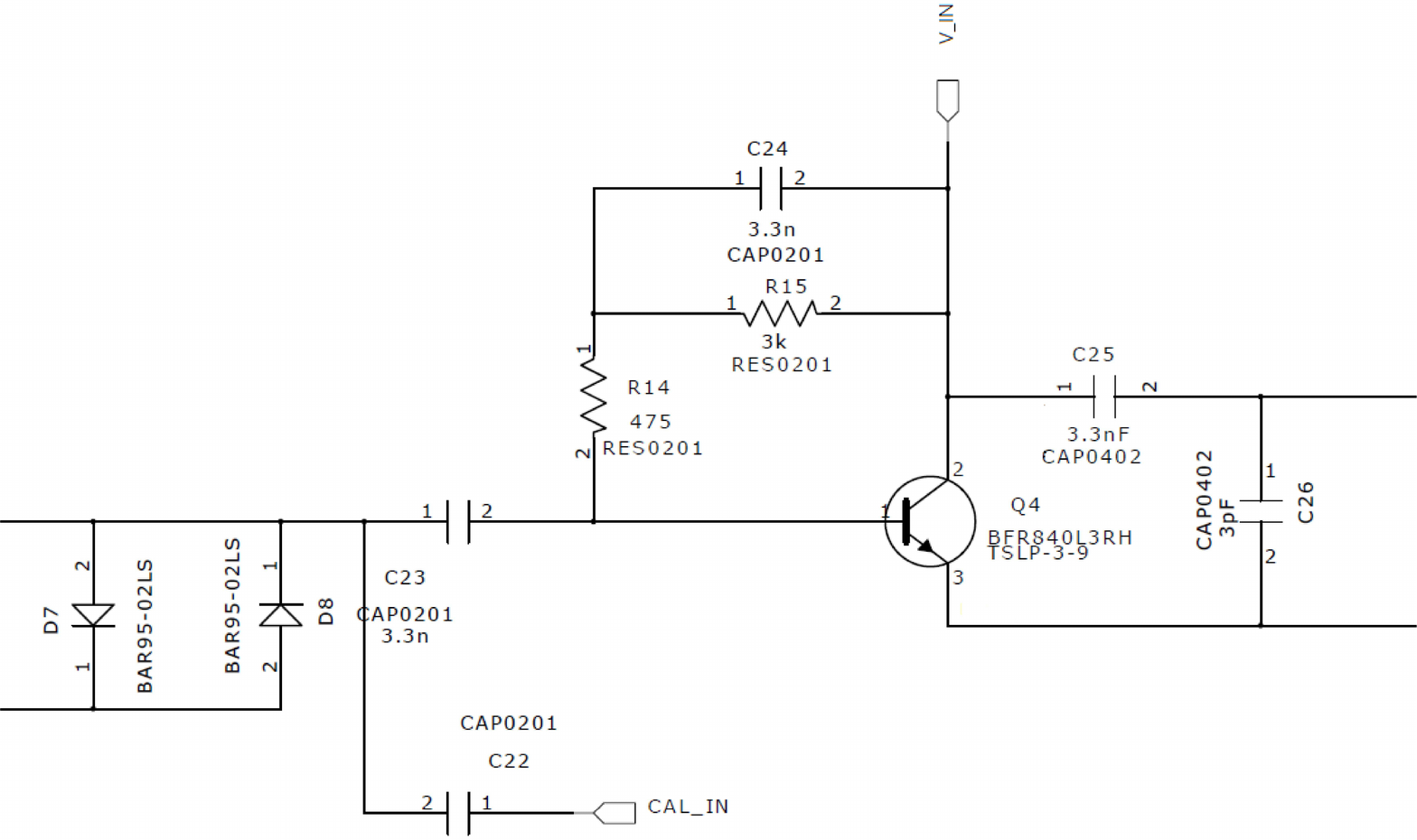}
    \label{fig:UCSC-amplifier}
  }
  \hfill
  \subfloat[]{
    \includegraphics[width=0.45\textwidth]{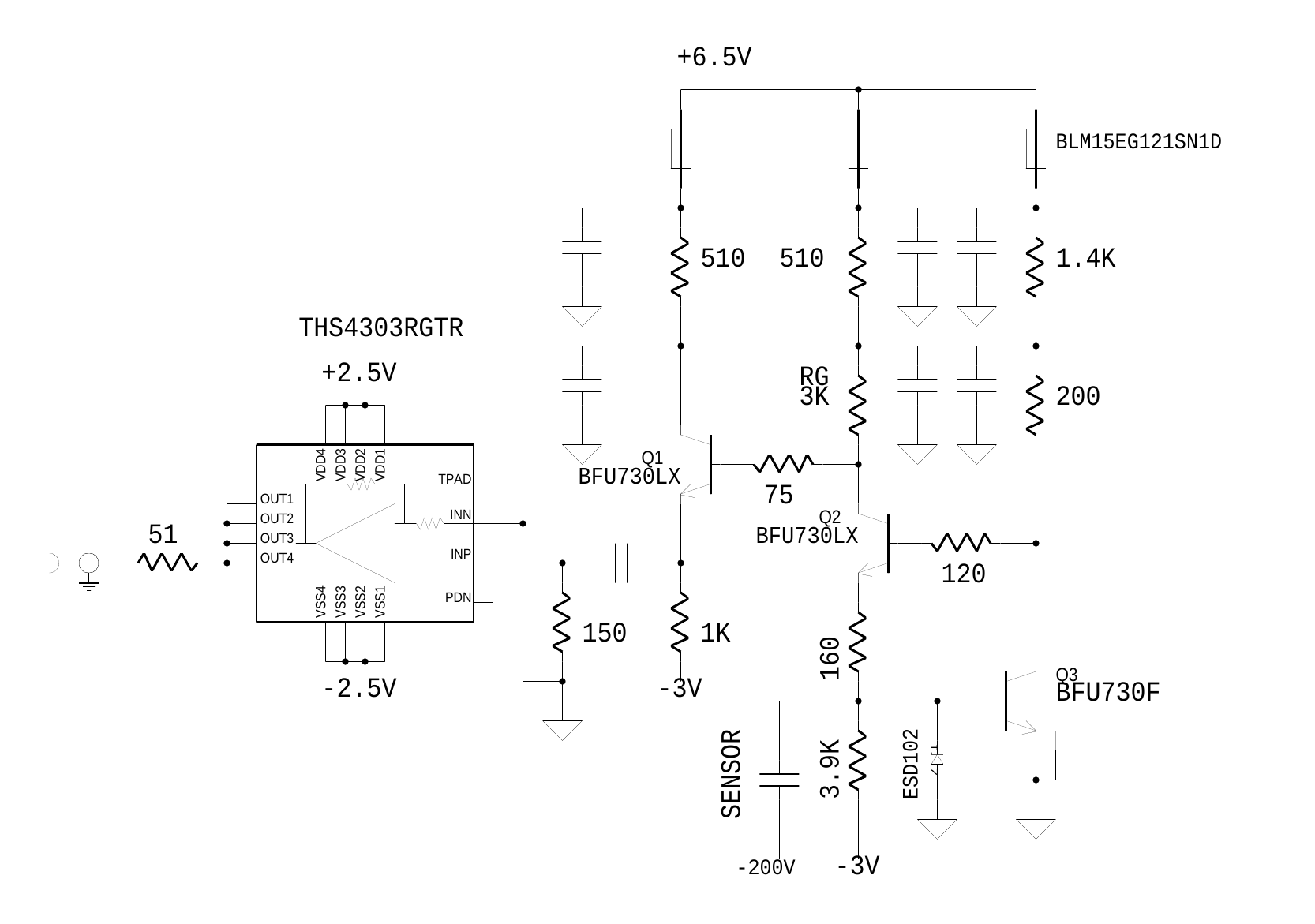}
    \label{fig:IN2P3-amplifier}
  }
\caption{Schematics of the first stage amplifier in boards \#1 and \#2 \protect\subref{fig:UCSC-amplifier} and of the amplifier of board \#3 \protect\subref{fig:IN2P3-amplifier}.}
\label{fig:amplifiers}
\end{figure}

Read-out board versions \#1 and \#2 were developed at the University of California Santa Cruz (UCSC)~\cite{bib:UFSD50umTBNicolo}.
Sensors are attached to the boards using double sided conductive tape while the amplifier input is coupled to the front side metallization layer
via multiple wire bonds to minimize inductance. A 1\,M$\Omega$ resistor attached between input and ground serves for detector biasing,
followed by a pair of low forward-resistance silicon pin diodes. The latter, with a 50\,V breakdown at 5~\SI{}{\micro\ampere},  functions as a protection
for the amplifier input.
The two board versions differ in the number of implemented amplification stages. In version \#1, the on-board amplifier includes only the
single stage, whose design is described hereafter, followed by a commercial external voltage amplifier\footnote{\url{https://ww2.minicircuits.com/pdfs/GALI-52+.pdf}} with a gain of about 10 and
a bandwidth of 2\,GHz. Board version \#2 incorporates three discrete amplification stages with a voltage divider between the second and the third,
resulting in a total gain of about 200 at a bandwidth of 1.6\,GHz.
The first stage, common to both board versions, is based on a single transistor common emitter design and acts as an inverting trans-impedance
amplifier. Simplified schematics are shown in Figure~\ref{fig:UCSC-amplifier}. Amplification is performed by an
AC coupled silicon-germanium bipolar transistor with a bandwidth of 75\,GHz. At a bandwidth of 1.9\,GHz a gain of 29\,dB is expected, with
an integrated output noise of 260~\SI{}{\micro\volt}. The feedback loop is designed for timing with small capacitance sensors inducing typical
rise times of the order of 800\,ps, with a feedback resistor of 470\,$\Omega$.
The overall trans-impedance of the two board designs within a 1.6\,GHz bandwidth and terminated into 50\,$\Omega$ is listed in Table~\ref{tab:deviceTable}.
These values include, in the case of board version \#1, also the external amplifier and are affected by an overall scale uncertainty of 10\%.  Care is
taken to provide complete hermetic shielding on both sides of the board up to a bandwidth of 3\,GHz, with RC filtering in both the high and low voltage
input lines. The PCB design has been optimized to minimize parasitics and reduce inductance on the signal return path using at least 6 decoupling high
voltage capacitors, 0201 size surface mount components and ground buried signal and power lines.

Read-out board version \#3 for 3$\times$3 mm$^2$ sensor arrays was developed by LAL/IN2P3.
Simplified schematics of its amplifier are shown in Figure~\ref{fig:IN2P3-amplifier}. The amplifier consists of a regulated cascode trans-impedance amplifier (Q2, Q3), an emitter follower (Q1)
and a voltage post-amplifier (THS4303 operational amplifier with a fixed gain of 10, 1.8\,GHz bandwidth). The front-end is realized with discrete components. The regulated cascode configuration is expected to achieve higher bandwidth with a relatively high-capacity detector (tens of pF) than would otherwise be possible with a larger input-impedance trans-impedance amplifier. The trans-impedance gain of the first stage is determined by RG=3\,k$\Omega$. The overall trans-impedance gain of the full amplifier, properly terminated into 50\,$\Omega$, is expected to be about 12.5\,k$\Omega$, within a 1\,GHz bandwidth. The circuit in its four-channel version was realized on a standard PCB (glass epoxy laminate) with the additional complication of providing adequate clearance for high voltage biasing of the LGAD sensor.  Precautions are taken to minimize parasitics: ground plane openings below the sensitive nodes; 0.4\,mm wide, short traces; usage of 0201-size based series resistors (to minimize inductance). However the initial version, with a higher Q3 bias current, was prone to oscillation at around 4\,GHz at the inner Q2-Q3 loop. The gain of the inner loop was therefore decreased; the actual biasing resistor values are those shown in the present schematics. A more detailed analysis indicates the instability is likely caused by parasitic inductance in that loop (especially in the Q2 base connection).
The LGAD sensor is connected to the PCB using a conductive glue: a technique developed by the CALICE collaboration\footnote{\url{https://twiki.cern.ch/twiki/bin/view/CALICE/WebHome}} and only applicable to large pads ($\ge$ 3$\times$3\,mm$^2$). A robot deposits some dots of glue
near the centre of each pad with an optical position control. The amount of glue (150\,\SI{}{\micro\meter} thickness) is calibrated to ensure a low-resistivity contact between the sensor pads and the PCB pads while avoiding any leakage to neighbouring pads. The positioning and
alignment of the sensors is crucial; the sensors should be placed within 100\,\SI{}{\micro\meter} of the lateral dimension of the PCB pads. Figure~\ref{fig:IN2P3-array-board} shows the PCB board with the glued sensor at the center.


\subsection{Cherenkov counters}
\label{sec:CherenkovCounter}
Fast Cherenkov counters with time resolutions similar to or better than the sensors~\cite{bib:SiPM,bib:AFPbeamTests} are used as timing references. They consist of Cherenkov-light emitting quartz bars of 10\,mm length along the beam, coupled to Silicon Photomultipliers (SiPM) of matching area. Details are given in Ref.~\cite{bib:UFSD50umTBNicolo}. To match the active area of the tested LGAD, one counter, referred to in the following as SiPM1, has an area of 3$\times$3\,mm${^2}$ transverse to the beam, and the other, referred to in the following as SiPM2, has an area of 6$\times$6\,mm${^2}$ transverse to the beam.

\subsection{Devices under test}

A complete list of the devices studied in this beam test is presented in Table \ref{tab:deviceTable}, including the sensor name, information on the pad multiplicity (single pads or arrays), size and pad capacitance, as well as on the implant dose of the p$^+$ multiplication layer and the total trans-impedance of the read-out board and external amplifier combination. Also the maximum voltage applied to each device is included, which was determined by a leakage current level of about 5\, \SI{}{\micro\ampere}, beyond which operation was not considered safe anymore due to reaching the breakdown regime.

\begin{table}
        \centering
        \caption{Devices measured in the beam test: LGAD single pads (``S'') and arrays (``A'') including information on the pad multiplicity, pad size, pad capacitance, implant dose of the p$^+$ multiplication layer, the total trans-impedance of the read-out board and external amplifier combination, and the maximum voltage applied. All sensors are 45\,\SI{}{\micro\meter} thick. }
        \label{tab:deviceTable}
        \begin{tabular}{|c|c|c|c|c|c|c|c|c|}
\hline
Device	&	Sensor	&	Pad 	&	Pad Size	&	C	&	p$^+$ Dose	&	Read-out 	&	Trans- & Max.	\\
 Name	&		&	 Mult.	&	 [mm$^2$]	&	 [pF]	&	 [$10^{13}$\,cm$^{-2}$]  	&	board v.	&	Imped. [$\Omega$] & V [V]	\\
\hline
S1M-1	&	W5 LGA31	&	1	&	1.3$\times$1.3	&	3.9	&	1.9	&	\#1	&	17540	& 240\\
S1M-2	&	W5 LGA33	&	1	&	1.3$\times$1.3	&	3.9	&	1.9	&	\#1	&	4700	& 240\\
S1H	&	W12 LGA34	&	1	&	1.3$\times$1.3	&	3.9	&	2.0	&	\#1	&	4700	& 85\\
A2M	&	W7 HG22	&	2$\times$2	&	2$\times$2	&	11	&	1.9	&	\#2	&	10700	& 180\\
A3M	&	W8 HG11	&	2$\times$2	&	3$\times$3	&	22	&	1.9	&	\#3	&	12500	& 200\\
\hline
\end{tabular}

\vspace{1cm}

\begin{tabular}{|c|c|c|c|}
\hline
\multicolumn{4}{|c|}{Nomenclature for Device Name and Read-out Board \#:} 	\\
\hline
Board	&	Pad size [mm]	&	p$^+$ Dose	&	\#	\\
\hline
``S'' = single	&	``1'' = 1.3	&	``L'' = low	&	for identical	\\
``A'' = 2x2 array	&	``2'' = 2	&	``M'' = medium	&	boards	\\
	&	``3'' = 3	&	``H'' = high	&		\\
\hline
\end{tabular}
\end{table}


\section{Beam test setup}
\label{sec:setup}

The results presented in the following are obtained from data collected in 2016 during a two-week beam test in October/November, at the H6B beam line of the CERN-SPS North Area with 120 GeV pions. Two data-taking modes can be distinguished: stand-alone and integrated into a beam telescope.

In stand-alone mode the pulses of up to 3 LGAD sensors were read out simultaneously by an Agilent Infiniium DSA91204A oscilloscope with 40\,GSample/s sampling rate and a bandwidth of 12\,GHz. Apart from single test runs, the bandwidth was mostly reduced to 2\,GHz for data taking, in order to reduce high frequency noise contributions. The vertical scale of the oscilloscope was adjusted for each run to only saturate the pulse height in a few percent of the events, while minimizing the quantification noise contribution from the oscilloscope. The Cherenkov counter (see Section~\ref{sec:CherenkovCounter}) with a time resolution expected to be significantly lower than that of the sensors under test was also included in the data taking, connected to the fourth and last available channel of the oscilloscope. Since the quartz bar had a much larger surface than the LGAD sensors, its use as a trigger would have made the geometrical efficiency very low. Therefore one of the sensors was used as a trigger, while voltage scans were performed on a different one. A custom-made support structure provided mechanical stability and the correct alignment of the sensors and SiPMs. The setup was mounted on a base plate connected to remotely controllable stage motors moving in the horizontal and vertical directions perpendicular to the beam direction with micrometer precision. This allowed for a precise positioning of the sensor at the centre of the beam. To provide light-tightness for the operation of the SiPM, the base plate was covered with a styrofoam box.

\begin{figure}[htbp]
\centering
  \subfloat[Sensors under test]{
    \includegraphics[width=0.5\textwidth]{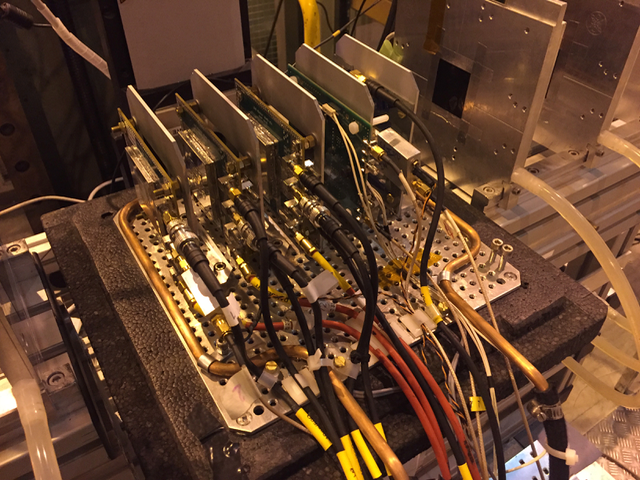}
    \label{fig:SetupPicture}
  }
  \hfill
  \subfloat[Data acquisition setup]{
    \includegraphics[width=0.45\textwidth]{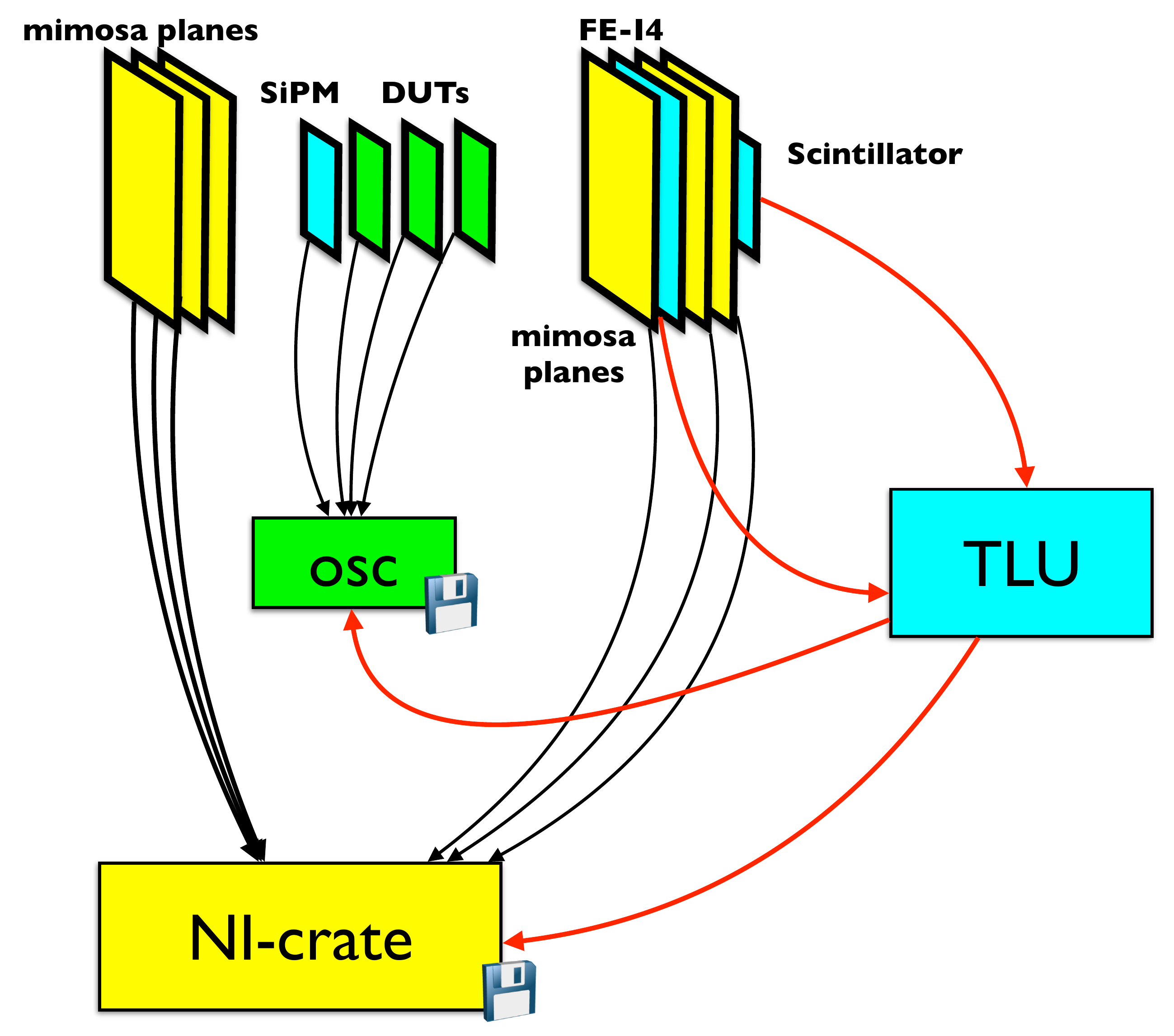}
    \label{fig:SetupDrawing}
  }
  \caption{Picture \protect\subref{fig:SetupPicture} of the sensors under test on the movable table and drawing \protect\subref{fig:SetupDrawing} of the beam test data acquisition setup}
  \label{fig:Setup}
\end{figure}

At a later stage a EUDET-type beam telescope based on MIMOSA pixel planes with a track position precision of few micrometers~\cite{Jansen:2016bkd} was also included in the data taking, allowing for a position-dependent measurement. A picture and a schematic drawing of the setup are shown in Figure \ref{fig:SetupDrawing}. The pulses of up to 8 sensors were read out by one or two oscilloscopes with 10 or 40\,GSample/s and 2 or 3\,GHz bandwidth. In the configurations with more than 4 channels, additionally to the same oscilloscope used in stand-alone mode, an Agilent Infiniium DSO9254A oscilloscope with a sampling rate of 20\,GSample/s for up to 2 channels (10\,GSample/s otherwise) and a bandwidth of 2.5\,GHz was used. Both oscilloscopes were set to use the same bandwidth and sampling rate.
Whenever the beam telescope was included in the data acquisition, the trigger was provided by the coincidence of signals on a scintillator and a special 3D FE-I4 plane.
This plane, consisting of a 3D CNM Silicon sensor connected to the FE-I4 read-out chip as also used in the ATLAS IBL~\cite{Aad:2012wf}, is a pixel detector with pixel size of 50 and 250\,\SI{}{\micro\meter} in the $x$ (horizontal) and $y$ (vertical) direction, respectively, and 25\,ns clock. It served two purposes. Firstly, it provided a so-called \emph{hitOr} trigger that fires when at least one of the pixels selected in a user-defined mask has a hit. Hence, it was used as a region-of-interest (ROI) trigger to only accept tracks traversing the small area of the LGAD sensors. Secondly, with its 25\,ns time binning of the hits it has a very short integration time, compared to the telescope planes that integrate hits over 112.5\,\SI{}{\micro\second} and hence usually provide multiple tracks within this period at the typical SPS particle rates. Hence, by matching the mostly unique signal of the FE-I4 plane to one of the several tracks provided by the telescope, the track that fired the trigger can be selected. The trigger signals were combined in the Trigger Logic Unit (TLU)~\cite{Jansen:2016bkd}, whose output was used by the telescope and connected to the oscilloscope, thus ensuring a perfect correspondence between the events recorded by the oscilloscope(s) and by the telescope. The two data acquisition chains were separate and the information from both systems were combined offline. The data from the LGAD sensors and Cherenkov counter were collected by the oscilloscope(s) described above, while the beam telescope and FE-I4 data were saved in a National Instrument (NI) PXIe crate ~\cite{Jansen:2016bkd}. The synchronicity of the two data streams was constantly monitored and never failed as long as the SPS provided a regular spill structure. Either the same support used for stand-alone data-taking or single metal frames (see Fig.~\ref{fig:SetupPicture}) were used to position the sensors perpendicular to the beam. The movable table, on which the sensors and SiPMs were positioned, was used to align the setup with the beam telescope and FE-I4 plane, while the whole system was aligned to the centre of the beam with another set of remotely controllable stage motors. The styrofoam box covering the devices under test was used both for light-tightness and to keep a stable temperature of about 25$^{\circ}$C with a water cooling system. A nitrogen outlet available in the area was inserted in the box to avoid humidity condensation that would damage the sensors.

\section{Data reconstruction and analysis methods}
\label{sec:reco}


\subsection{Oscilloscope data reconstruction}
For each event and for each channel, at least 2000 samples were registered spaced every 25\,ps. The first step in the oscilloscope data reconstruction is to measure the pedestal and the noise computed as the mean and standard deviation of the measured voltage, respectively, using the first 240 samples where no signal contribution is expected. The pedestal varies from 1 to 5\,mV depending on the run conditions and oscilloscope settings and is subtracted from the measured pulses event-by-event. The noise was found to be independent of the bias voltage, as expected, since the leakage current of the sensor is negligible with respect to the electronic noise. The run-by-run variations (up to 20\%) are attributed to different settings of the oscilloscope. Table~\ref{tab:noise} shows the average noise measured for each sensor. For arrays the noise correlation between pads was also measured and found to be negligible for A2M and of the order of 30\% for A3M, due to the different read-out board versions.

\begin{table}
  \small
  \begin{center}
    \caption{\label{tab:noise} Noise for all sensors. The statistical uncertainties are below 0.006 mV.}
    \begin{tabular}{|c|c|c|c|c|}
      \hline
     Sensor & \multicolumn{4}{c|}{Noise (mV)}\\
      \hline
     & \multicolumn{4}{c|}{Single pad sensors}\\
      \hline
     S1M-1 & \multicolumn{4}{c|}{2.6}\\
           \hline
     S1M-2 & \multicolumn{4}{c|}{2.2}\\
           \hline
     S1H & \multicolumn{4}{c|}{2.2}\\
      \hline
     & \multicolumn{4}{c|}{Arrays - pad number}\\
     & 1 & 2 & 3 & 4\\
           \hline
A2M &4.6 &5.4 &4.9  &4.7\\
           \hline
A3M  & 2.3& 2.4 &2.7 &2.3\\
      \hline
    \end{tabular}
  \end{center}
\end{table}



The maximum of the pulse amplitude is estimated after pedestal subtraction with a second-degree polynomial fit around the sample with the highest amplitude in a 400\,ps window. The collected charge is defined as the integral of the signal voltage after pedestal subtraction divided by the trans-impedance (see Table~\ref{tab:deviceTable}). The integral is computed numerically in a window centered around the time where the pulse is maximal and wide enough to fully contain the pulse. The gain is then obtained by dividing the collected charge by the expected charge from a MIP in a silicon sensor without gain ($Q^{\mathrm{no\:gain}}$). For a 45\,\SI{}{\micro\meter} thick sensor, the value of $Q^{\mathrm{no\:gain}}$ is 0.46\,fC~\cite{bib:EnergyLossSi}. The gain has an estimated systematic error of 20\% due to the uncertainty on the trans-impedance. 

Selections on the maximum amplitude are applied to reject noisy events (lower cut) and to reject saturated pulses that might be caused by the oscilloscope or the read-out electronics (upper cut). These selections are derived separately for each channel and each run. 

The optimal time resolution is expected to be reached with sophisticated time reconstruction techniques using the full information of the pulse shape (e.g.~digital filtering). However, since these techniques require a too large data bandwidth and cannot be used for the HGTD read-out, only three time reconstruction algorithms using discriminators are investigated in this paper. The first and simplest one is the Constant Threshold Discriminator (CTD) method where the time of arrival is defined as the time where the signal crosses a constant threshold. In order to be above the noise level, a value of 20\,mV has been chosen. The time of arrival is determined from a linear interpolation between the samples just above and below the threshold.
The drawback of the CTD method is that it suffers from time walk effects, which can be corrected if the signal amplitude is known.
Quite often the amplitude is not directly measured and the Time-Over-Threshold (TOT) information, correlated with the amplitude, is provided by the discriminator output.
The Constant Fraction Discriminator (CFD) method minimises the time walk effect by defining the time of arrival as the time where the signal crosses a constant fraction ($f_{CFD}$) of the maximum amplitude.
However, since the threshold is crossed before the maximum amplitude is reached, this method cannot be implemented in the read-out electronics. A third method reconstructs a time of arrival that is independent of the amplitude of the signal: the Zero-Crossing Discriminator (ZCD). A copy of the signal is delayed by $d_{ZCD}$ and attenuated by a factor $f_{ZCD}$. The zero-crossing time of the difference of the original signal and the attenuated copy is by construction independent of the signal amplitude under the assumption that the pulse shape remains identical. The CTD method with TOT-based time walk correction and the ZCD\footnote{One significant difference between the ZCD method implemented in this paper and the one under investigation for the HGTD front-end electronics is the use of an additional arming discriminator necessary to gate the zero-crossing discriminator output and avoid triggers induced by the noise which is not mandatory for an offline analysis.} methods are the two methods under investigation for the HGTD front-end electronics.

The time of the SiPM is always reconstructed using the CFD method with $f_{CFD}=0.2$. The optimization of the CFD and ZCD algorithms for LGADs is presented in Section~\ref{sec:opti}.
The default time reconstruction algorithm for LGAD sensors used in the following is the ZCD method.

The time resolution can be extracted from the width of the time differences computed from the LGADs and the SiPMs. Assuming that N devices with time resolutions $\sigma_{k}$ are used, there are N$\cdot$(N-1)/2 possible combinations. Assuming that the time resolutions of the devices are independent,  for each combination, one has:
\begin{equation}
\sigma_{ij}=\sigma_{i}\oplus \sigma_{j} 
\end{equation}
where $\sigma_{ij}$ is the width of the time difference distribution between device $i$ and $j$ and it is estimated as the width of a Gaussian function fitted iteratively on a range [-3$\sigma_{ij}$,3$\sigma_{ij}$].
Therefore, one has N unknowns and N$\cdot$(N-1)/2 constraints. For N$=$2, the system is under-constrained and no solution can be found without further assumptions. For N$=$3, the number of constraints equals the number of unknowns. In this case, the system of equations is linear considering the square of the time resolution as the unknown and it can be solved analytically. For N$>$3, the system is over-constrained and in order to fully use the available information, the time resolution can be extracted using a $\chi^2$ minimization technique:

\begin{equation}
\chi^2 = \sum_{i=1}^{N}\sum_{j=1}^{j<i} \frac{(\sigma^{2}_{ij}-\sigma^{2}_{i} - \sigma^{2}_{j})^{2}}{\sigma_{\sigma_{ij}^{2}}^{2}} 
\end{equation}
where $\sigma_{\sigma_{ij}^{2}}$ is the uncertainty on $\sigma_{ij}^{2}$.
The time resolutions are extracted using the $\chi^{2}$ method for runs where 4 measurements are available (3 LGADs and 1 SiPM). The time resolutions of SiPM1 and SiPM2 are found to be (10.9$\pm$0.8)\,ps and (35.3$\pm$1.9)\,ps, respectively, with up to 10\% variation due to varying running conditions (e.g.~bias voltage settings). The time resolution measurements of the LGADs are presented in Section~\ref{sec:TimeRes}.

\label{sec:TimeReco}

\subsection{Telescope data reconstruction and performance}
The beam telescope consists of  6 MIMOSA planes.
Planes 0, 1 and 2 are located upstream of the DUTs and planes 3, 4 and 5 are located downstream of the DUTs.  The  FE-I4 plane used
 for triggering  is located between the MIMOSA planes 3 and 4. The setup is shown in Figure~\ref{fig:SetupDrawing}. The positions of
 the telescope, FE-I4 and DUT planes are known with a precision of  1~mm in the $z$ direction along the beam line.
The dimension of the MIMOSA planes is 10.6$\times$21.2 mm$^2$ in the $x$ and $y$ directions with a pixel size of 18.5$\times$18.5\,\SI{}{\micro\meter}$^2$.
In order to reconstruct the hit position on each DUT plane, tracks are reconstructed using the information from all the 6 planes  
(or only 4 of them in a small part of the dataset where 2 of the planes had readout problems).

The first step in the reconstruction of tracks is the removal of the ``hot'' pixels  from the MIMOSA planes. Clusters are then built  from the  
remaining hits in each plane. 
Only clusters with a maximum of 6 hits are used for tracking. In the FE-I4 plane, a cluster has a maximum  of  2 neighbouring hits.
The cluster coordinates are the mean values of the hit coordinates in $x$ and $y$. In order to select events with only one particle traversing the DUTs, only events with 
exactly 1 such cluster in the FE-I4 plane are considered (about 94~\% of the total number of events).

The MIMOSA planes are aligned by iteratively
shifting the planes coordinates in $x$ and $y$ with respect to a reference plane, in order to minimize 
the difference between the reconstructed track position at
the MIMOSA plane and the measured hit position in the same plane.  

\begin{figure}[htbp]
\centering
   \begin{tabular}{ll}
   \hfill
   \subfloat[]{   
   \includegraphics[width=0.40\textwidth]{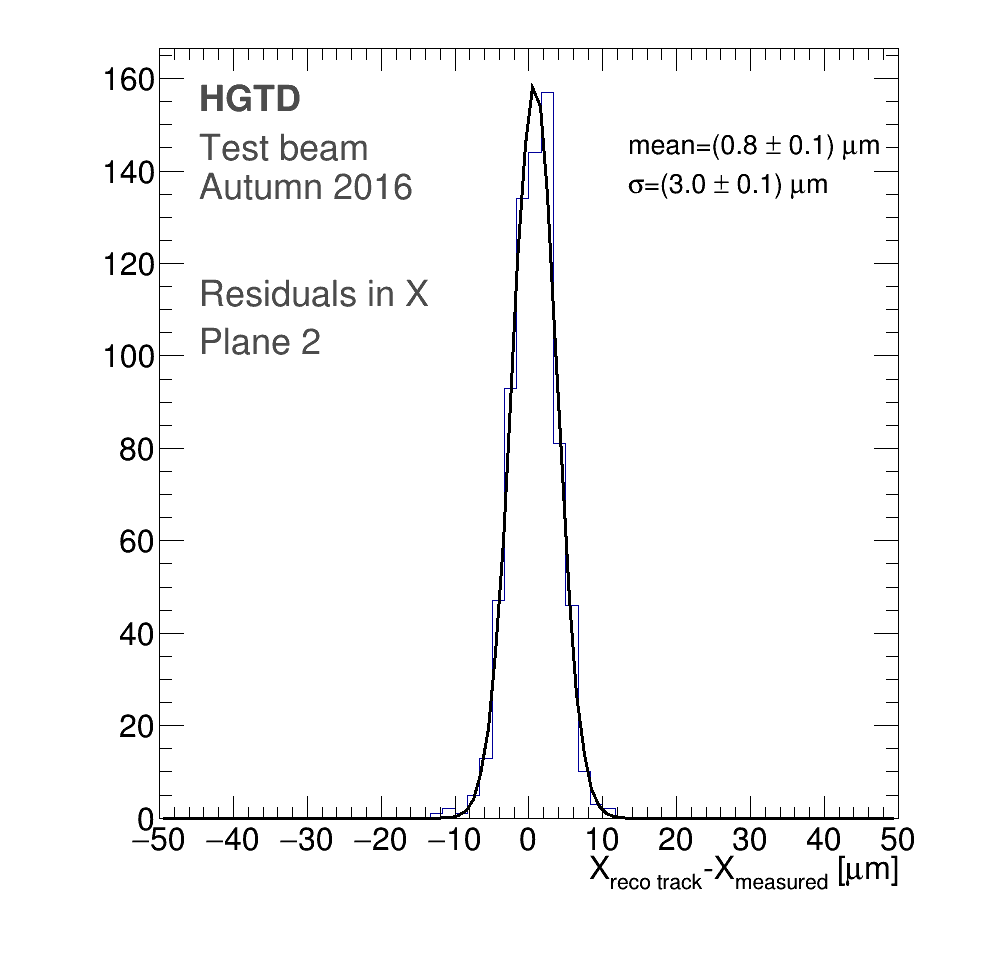}
   \label{fig:resplane2x}
  }
  \hfill  
   \subfloat[]{   
   \includegraphics[width=0.40\textwidth]{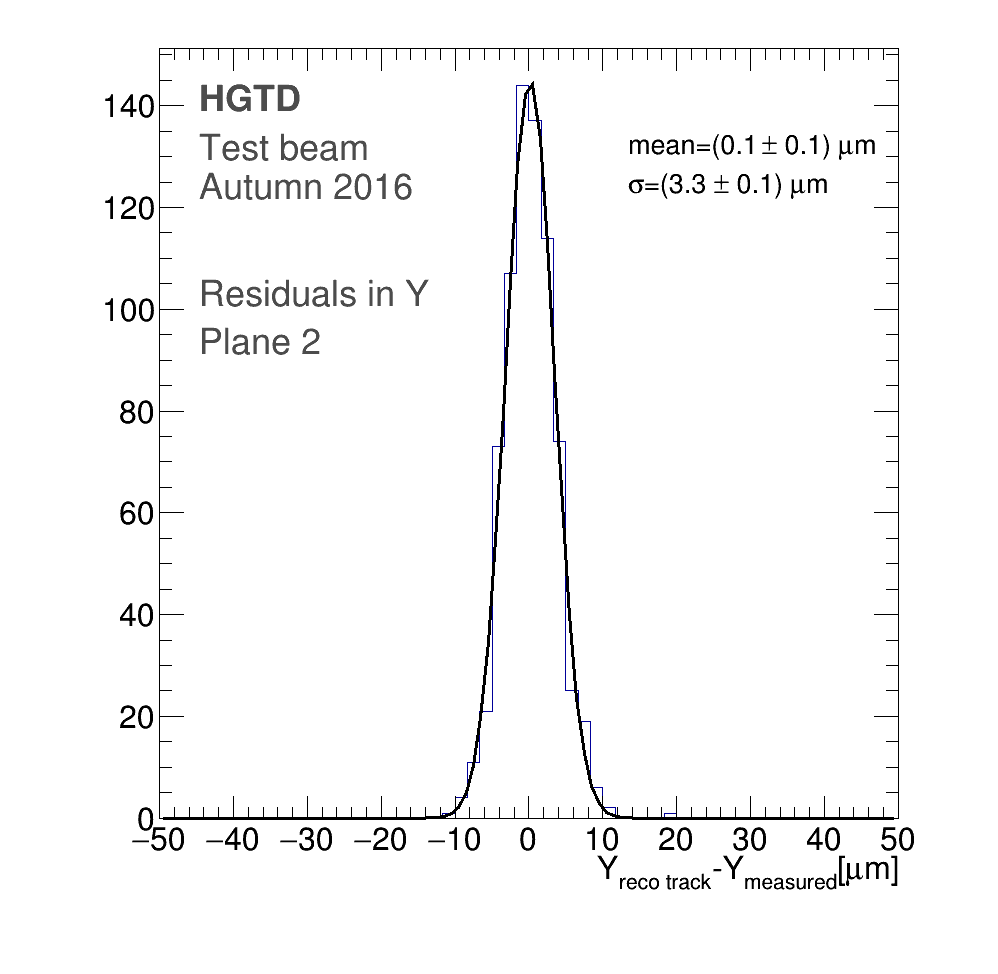}
   \label{fig:resplane2y}
  } \\
 \end{tabular}
  \caption{The residuals after the tracking procedure in the MIMOSA  plane 2  for  the horizontal \protect\subref{fig:resplane2x} and  vertical \protect\subref{fig:resplane2y} directions.  }
   \label{fig:residuals2}  
\end{figure}

Once the planes are aligned, the track fitting procedure is applied: knowing the $z$-position of the MIMOSA planes along
the beam axis and the ($x$,$y$) positions of the hits in these planes, 3D-tracks are built from the six planes of the 
telescope starting with the planes closest to the FE-I4. 
The reconstructed tracks must coincide  with a hit in the FE-I4 plane 
and  only events with a single reconstructed track through the six MIMOSA planes are considered.  
The residuals, defined as the difference between the track position in a MIMOSA plane and the hit used to build the track, are 
shown in Figure~\ref{fig:residuals2} for plane 2 (the closest to the DUT's). 
The precision on the position of the extrapolated reconstructed track in the DUT planes is about 3\,\SI{}{\micro\meter} 
in the $x$ and $y$ directions.

The tracking efficiency, defined as the number of events with one
reconstructed track divided by the total number of events with one
cluster in the FE-I4 is about 77\% mostly due to the exclusion of
inefficient and noisy regions in the MIMOSA planes. Furthermore only
straight tracks were considered in order to have a precise
interpolation to the DUT.


\label{sec:TeleReco}

\section{Results}
\label{sec:results}

\subsection{Pulse properties}
\begin{figure}[htbp]
  \centering
  \subfloat[]{
    \includegraphics[width=0.45\textwidth]{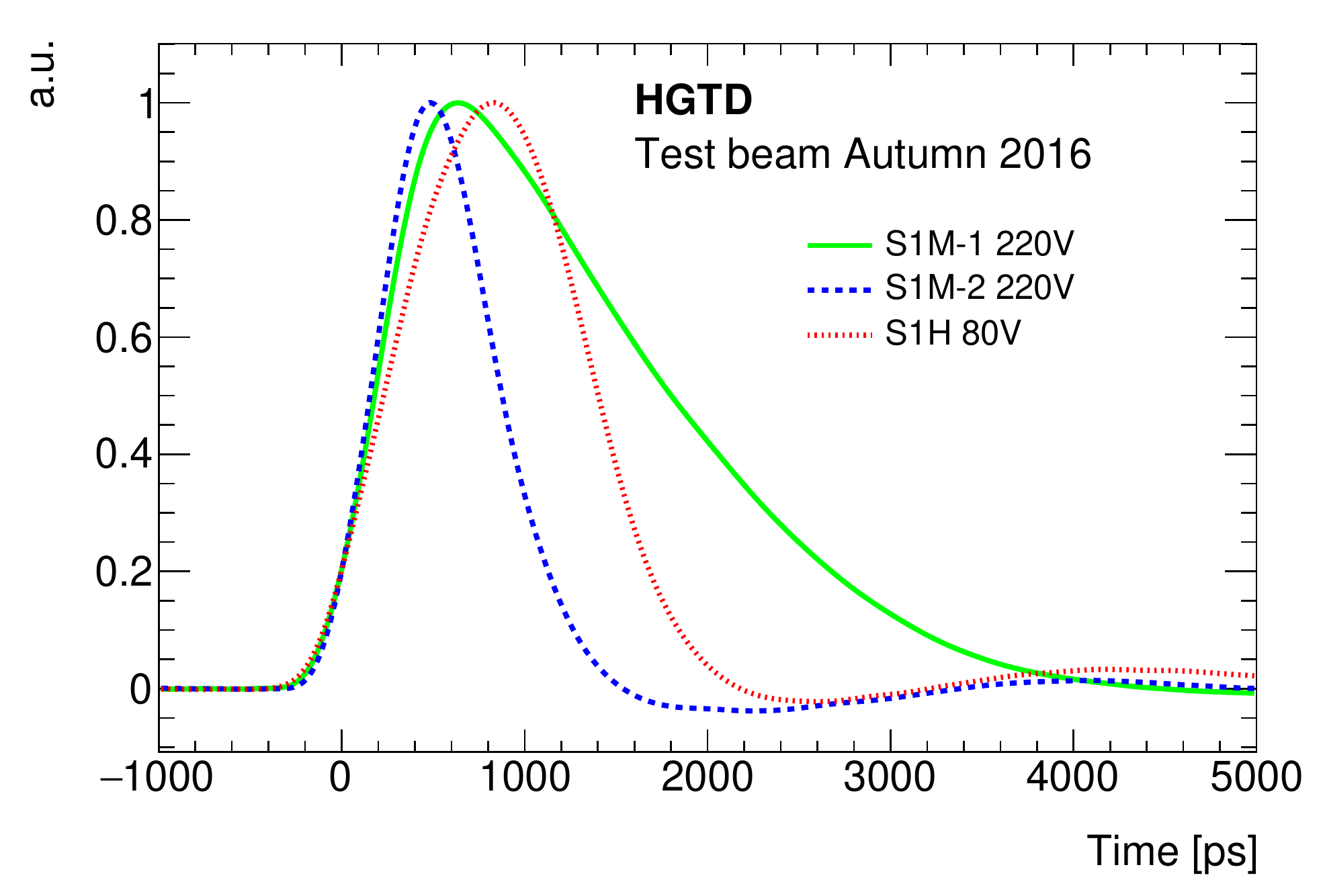}
    \label{fig:SPWidth}
  }
  \hfill
  \subfloat[]{
    \includegraphics[width=0.45\textwidth]{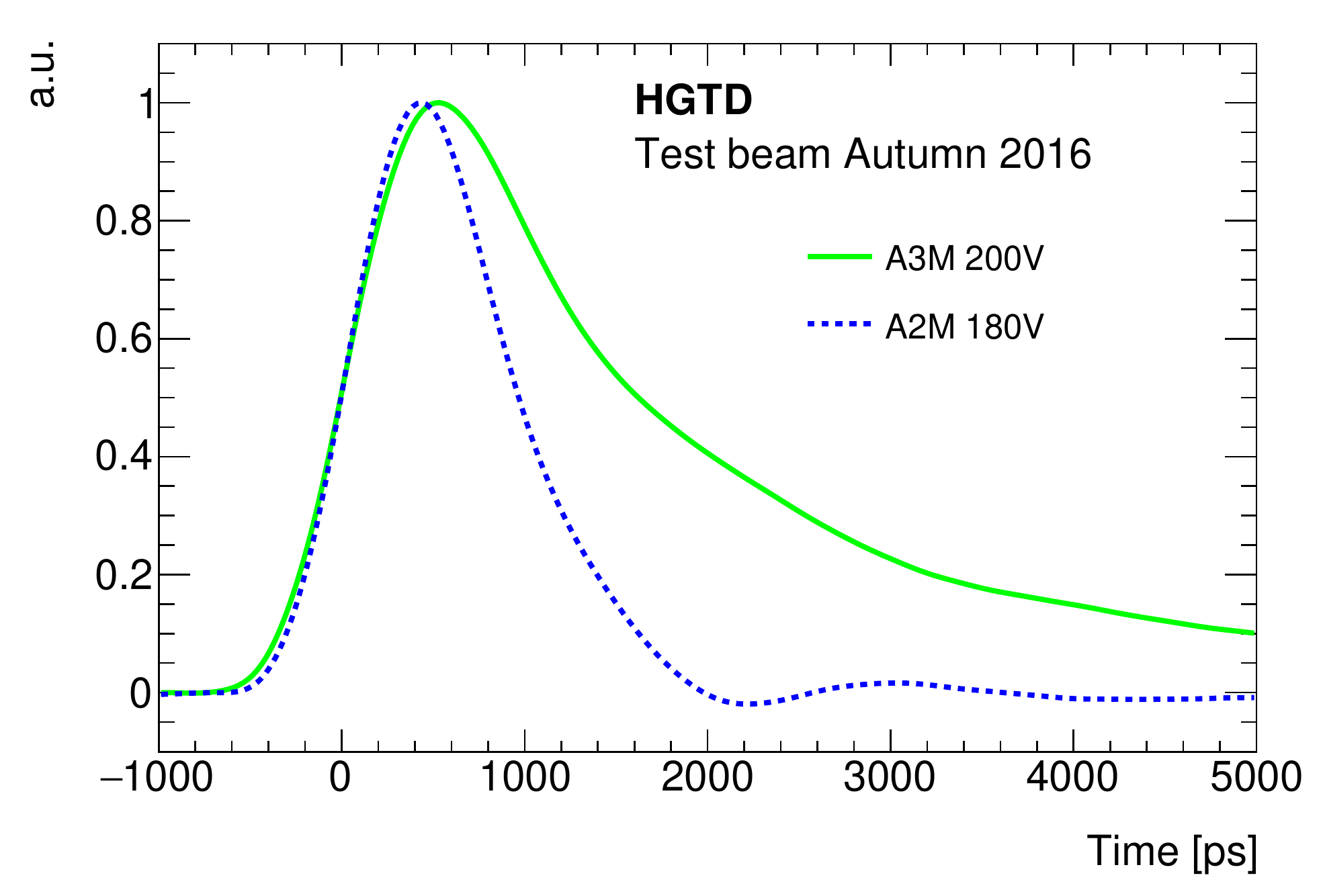}
    \label{fig:AWidth}
  }
  \caption{Averaged pulse shapes for single pad sensors \protect\subref{fig:SPWidth} and for pad 1 of the  arrays \protect\subref{fig:AWidth}.}
  \label{fig:PulseShapes}
\end{figure}

The performance of a LGAD sensor depends strongly on the characteristics of
its pulse shape (e.g.~charge and rise time). Figure~\ref{fig:PulseShapes} shows
 the averaged pulse shapes for single pads and arrays normalized to the maximum pulse amplitude.
The averaging was performed after synchronising the signals.
The pulse shapes are given by the convolution of the intrinsic LGAD waveform
with the electronics response functions, therefore they are different for the
 identical sensors, S1M-1 and S1M-2, read out by boards with different trans-impedance amplifier characteristics. The pulse shapes of S1M-2 at 220 V and S1H at 80 V, differ slightly, in spite of the sensors being connected to the same amplifier type, due to the significantly lower bias voltage and hence drift velocity for S1H, which leads to longer rise time and larger pulse width.

\subsubsection{Amplitude and charge}
\begin{figure}[htbp]
  \centering
  \subfloat[]{
    \includegraphics[width=0.45\textwidth]{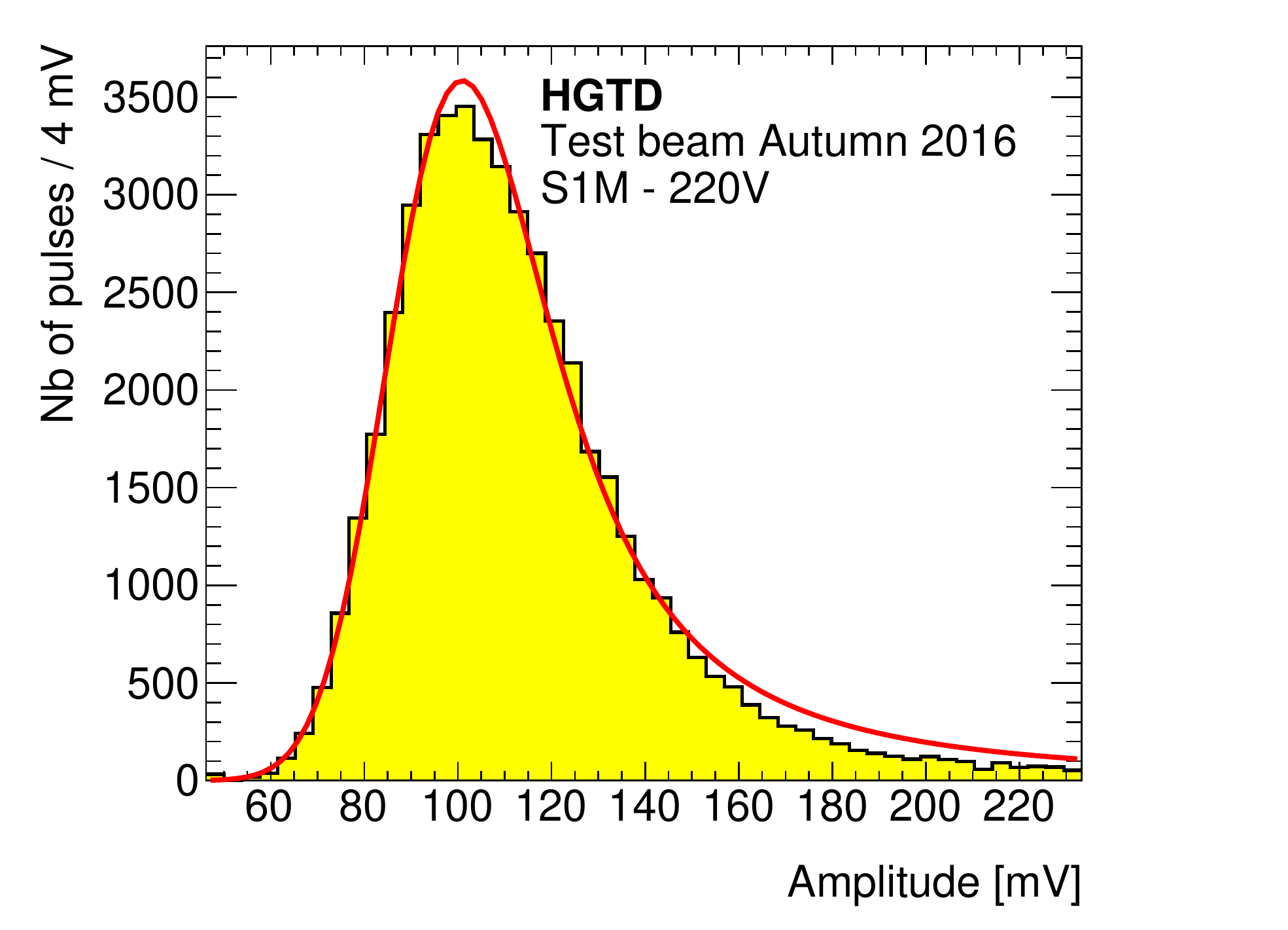}
    \label{fig:SPAmplitude}
  }
  \hfill
  \subfloat[]{
    \includegraphics[width=0.45\textwidth]{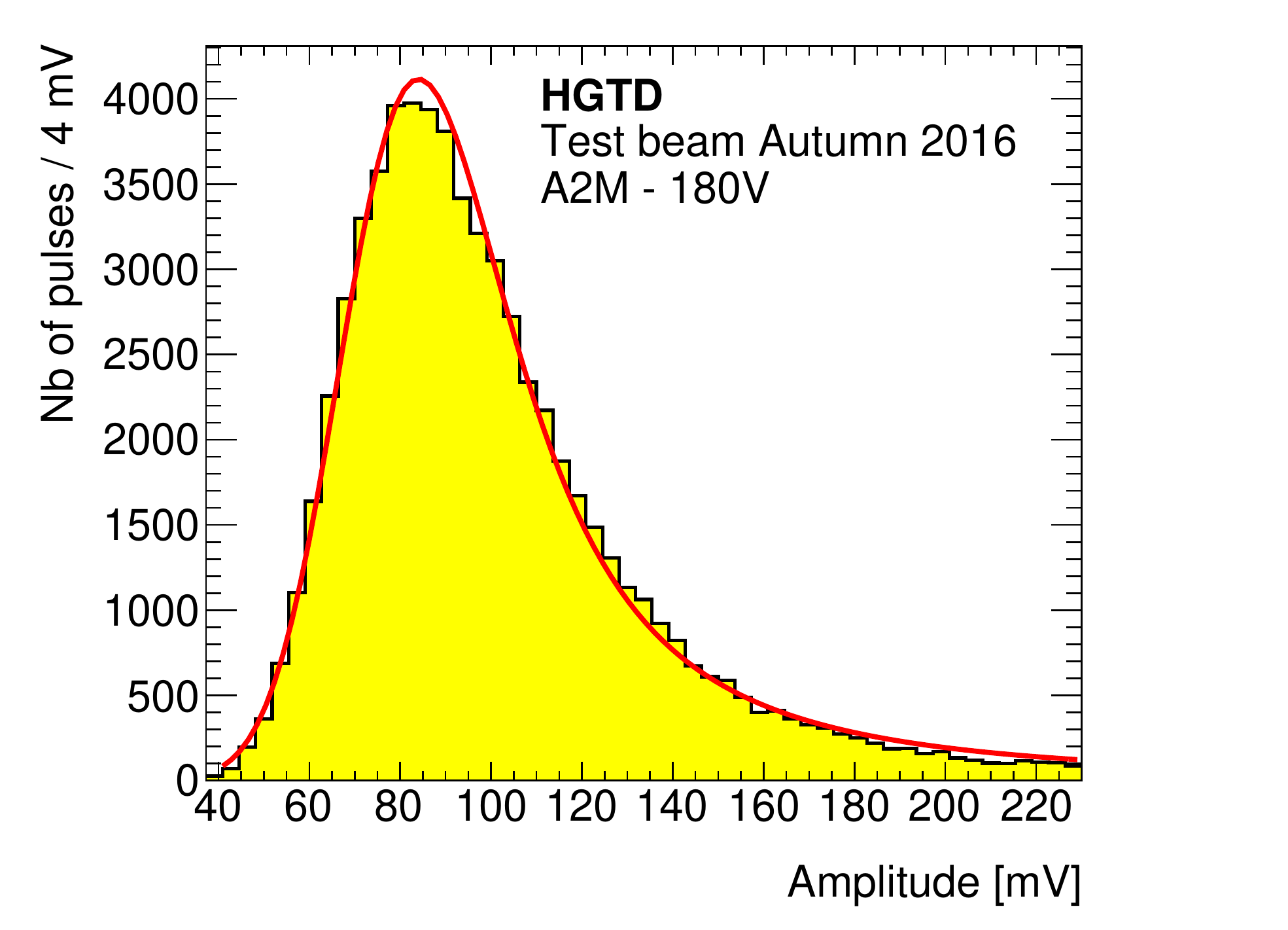}
    \label{fig:AAmplitude}
  }
  \caption{Distributions of the reconstructed maximum amplitude for the single pad S1M-2 \protect\subref{fig:SPAmplitude} and for one pad of the array A2M \protect\subref{fig:AAmplitude} fitted with a Landau function convoluted with a Gaussian function (red lines).}
  \label{fig:Amplitude}
\end{figure}

\begin{figure}[htbp]
  \centering
  \subfloat[]{
    \includegraphics[width=0.45\textwidth]{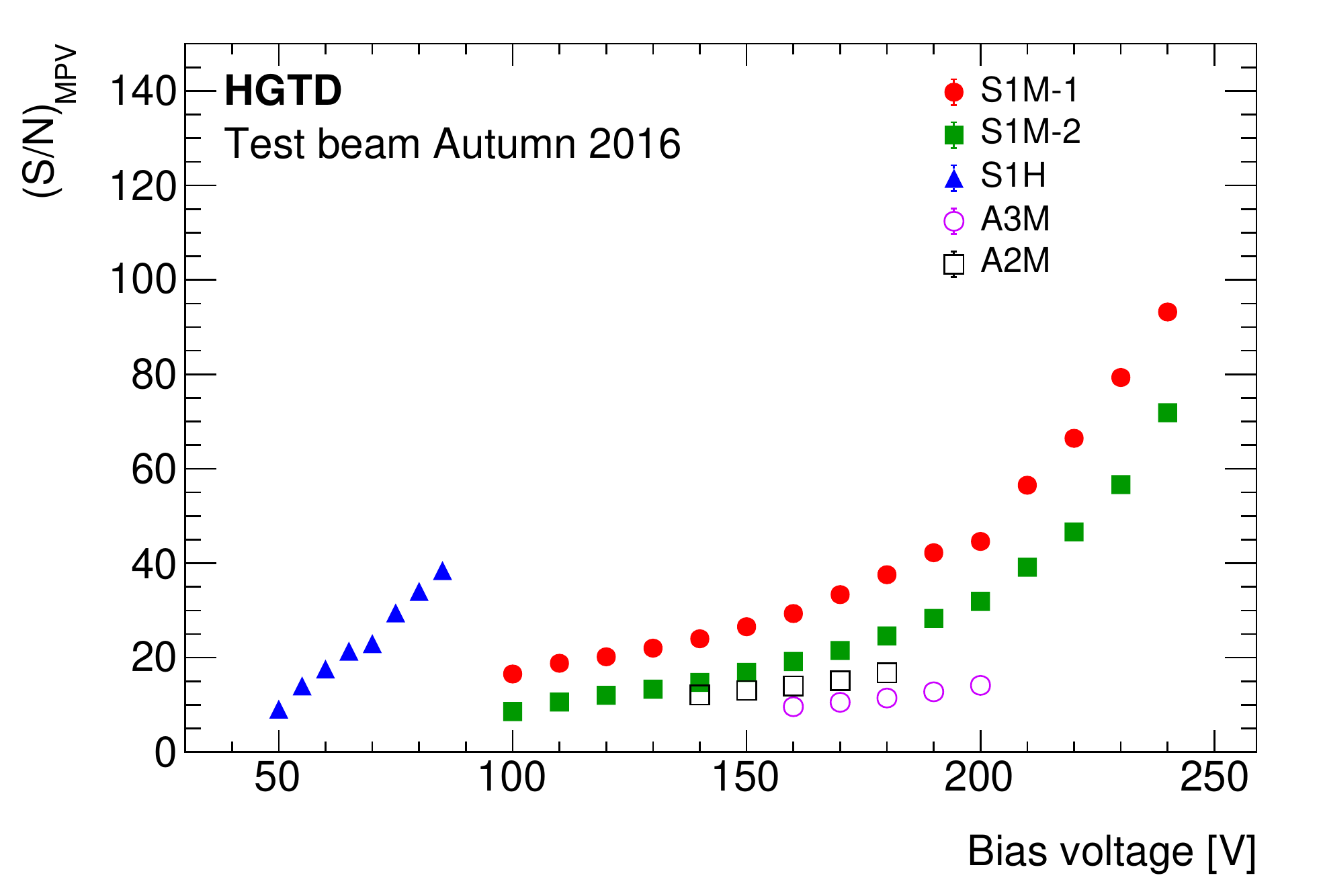}
    \label{fig:SOverNScan}
  }
  \hfill
  \subfloat[]{
    \includegraphics[width=0.45\textwidth]{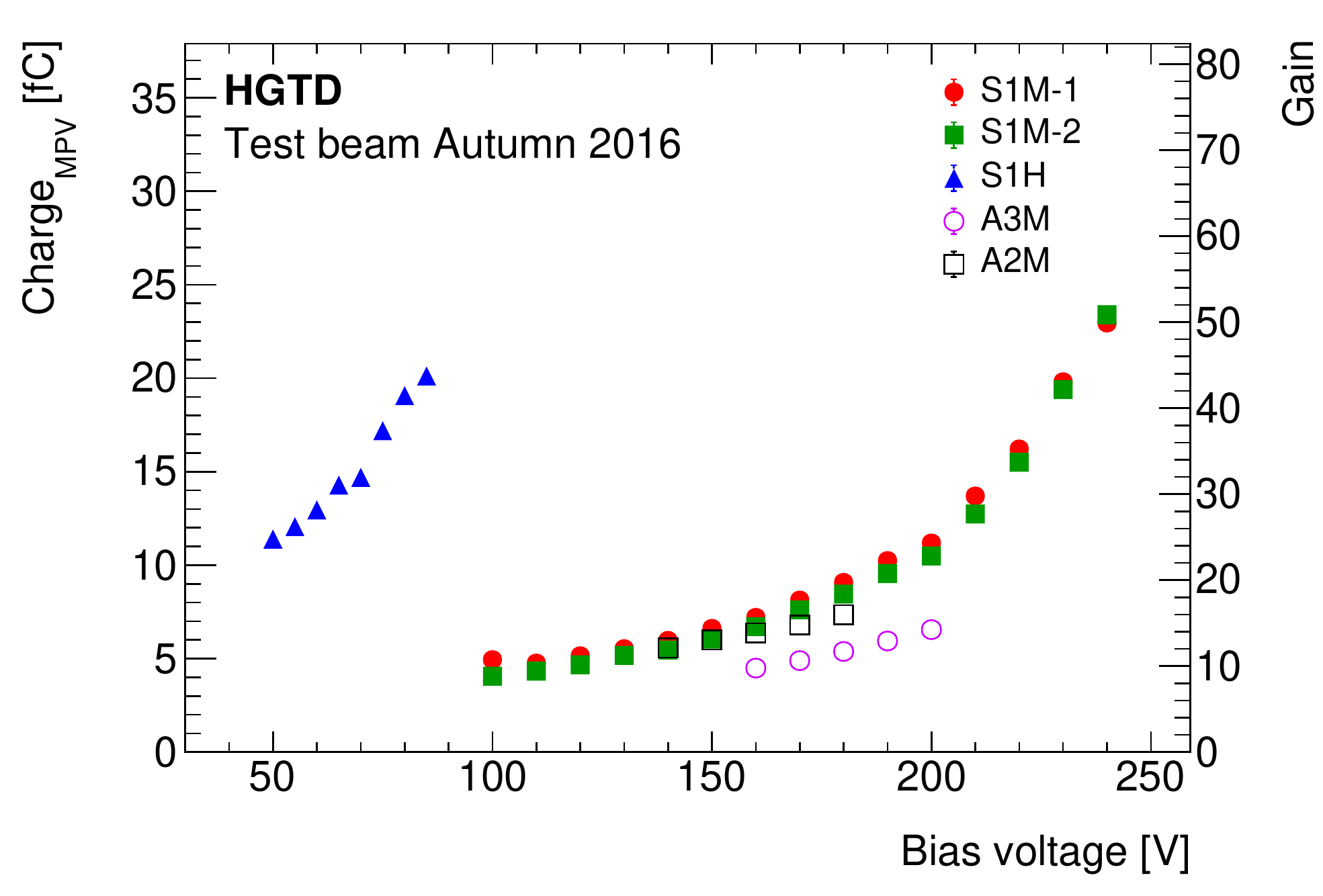}
    \label{fig:ChargeScan}
  }
  \caption{Signal-to-noise ratio \protect\subref{fig:SOverNScan} and charge and gain \protect\subref{fig:ChargeScan} as a function of the bias voltage for single-pad sensors and arrays.  Statistical uncertainties are negligible and smaller than the marker size.}
\end{figure}

Examples of the maximum pulse amplitude distributions are shown in Figure~\ref{fig:Amplitude}.
In order to estimate the most probable value, the distributions are fitted
 with a Landau function
convoluted with a Gaussian function.
These most probable values are used to compute the signal-to-noise
 ratio (S/N), that is shown in Figure~\ref{fig:SOverNScan} as a function of the bias voltage.
As expected, S/N increases with increasing bias voltage. Due to their larger
 capacitance, arrays have
lower signal-to-noise ratios compared to single pads. Larger signal-to-noise
 ratio is measured for S1M-1 compared to S1M-2 due to the larger
 trans-impedance of its read-out board.
The best signal-to-noise ratio is reached for S1M-1 at 240V with a value of 94.

The charge distributions are fitted with the same function used for the amplitude in order to extract the most probable value.
Figure~\ref{fig:ChargeScan} shows the most probable value of the charge
and the gain as a function of the bias voltage. Sensors with medium doping
need higher bias voltage to reach the same gain as sensors with high doping.
For instance, a gain of 30 is reached at 210~V for sensors with medium doping compared to 65~V for sensors with high doping.
As expected, the gain is similar for S1M-1 and S1M-2 and the highest gain
($\sim$50) is obtained for the largest bias voltage. The gain is found to be smaller for the arrays with medium doping and the highest value ($\sim$16) is obtained for A2M. Sensors of the same thickness and with the same doping are expected, within uncertainties, to have the same gain for a given bias voltage.
However, the sensor A3M equipped with a quite different preamplifier shows a smaller gain than expected. The most probable cause is the trans-impedance being incorrectly determined for the used bandwidth.

\begin{figure}[htbp]
  \centering
    \includegraphics[width=0.55\textwidth]{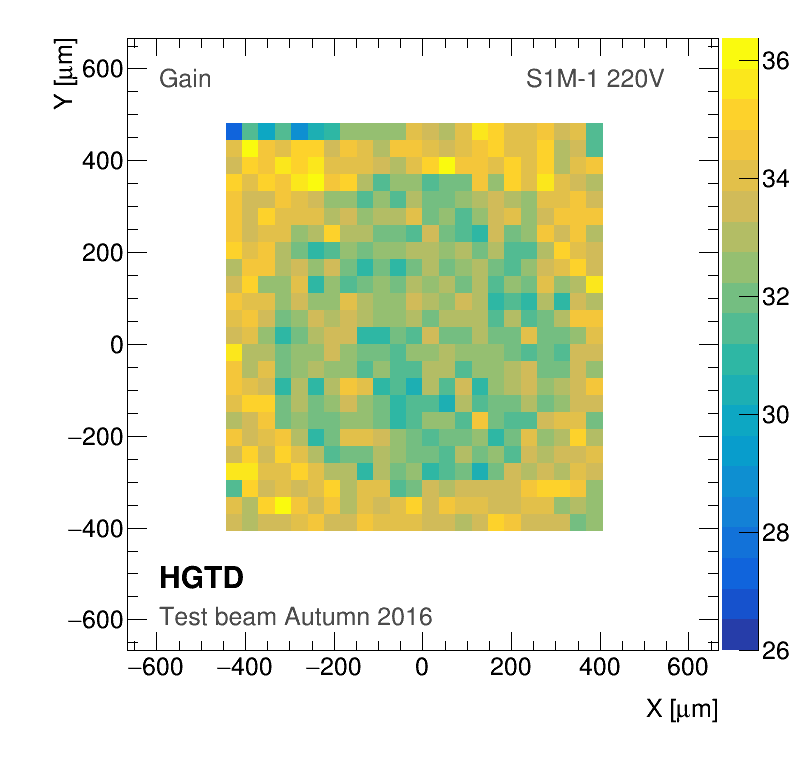}
    \label{fig:gain31_Pos}
  \caption{Gain for the single pad sensor S1M-1 as a function of the position in the pad with a bias voltage of  220\,V.
Each bin of size $(37~\SI{}{\micro\meter})^2 $ contains at least 40 events. 
}
  \label{fig:Sgain_vs_pos}
\end{figure}
\begin{figure}[htbp]
  \centering
  \subfloat[]{
    \includegraphics[width=0.45\textwidth]{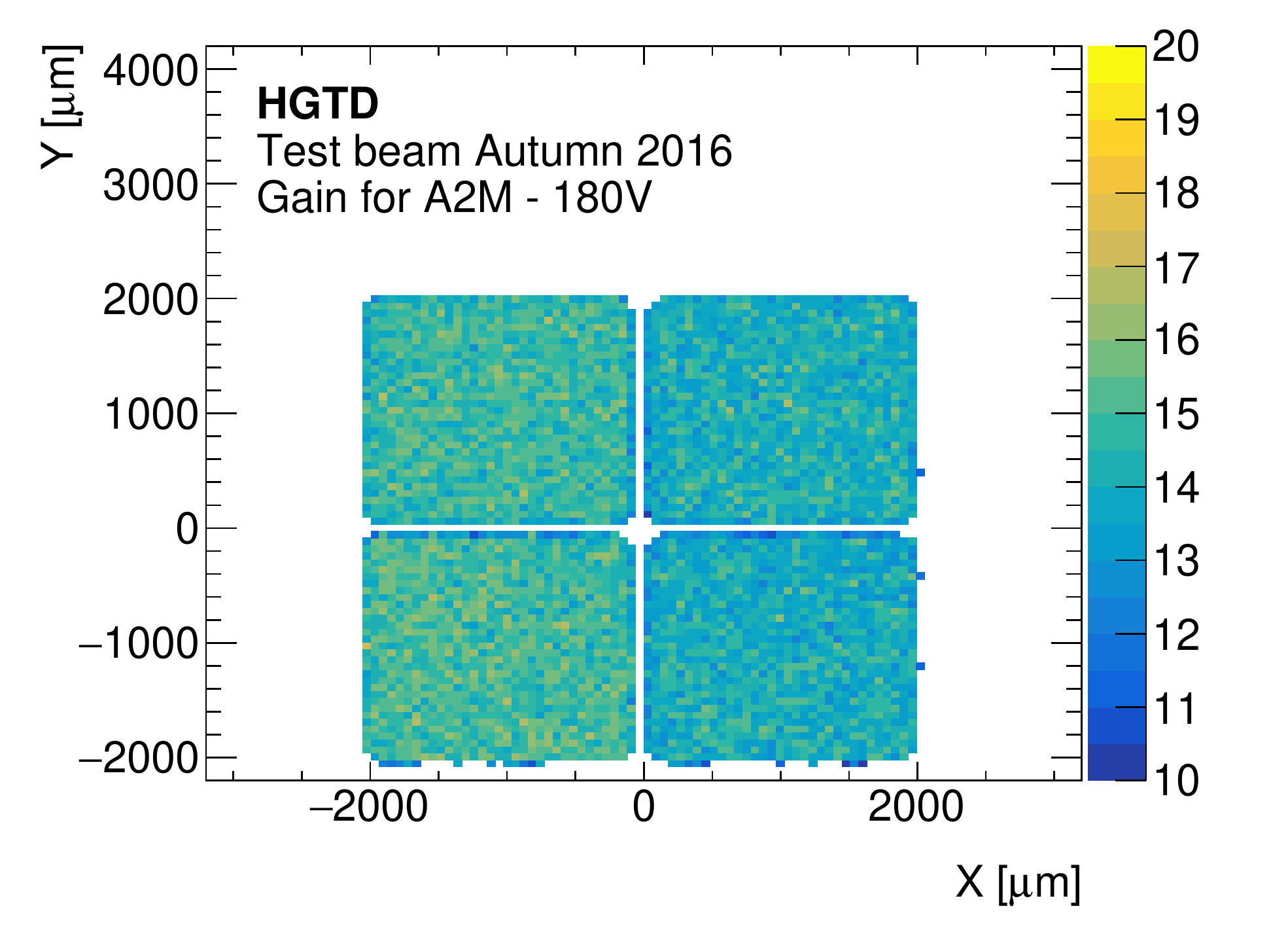}
    \label{fig:DiffGainA2M}
  }
  \hfill
  \subfloat[]{
    \includegraphics[width=0.45\textwidth]{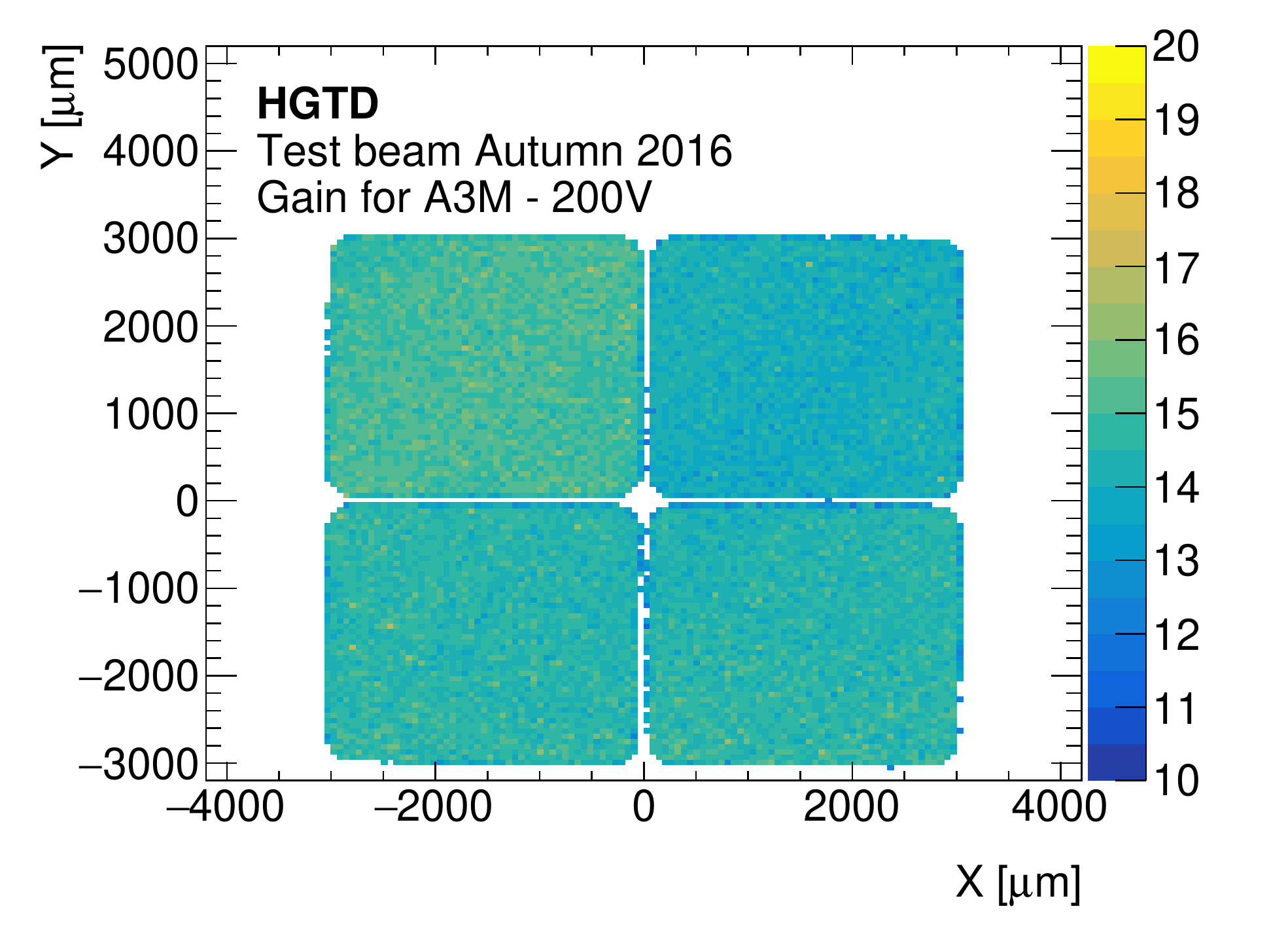}
    \label{fig:DiffGainA3M}
  }
  \caption{Gain for the array sensors A2M \protect\subref{fig:DiffGainA2M} and A3M \protect\subref{fig:DiffGainA3M} as a function of the position on the sensor.
Each bin of size $(60~\SI{}{\micro\meter})^2 $ contains at least 60 events.}
  \label{fig:gain_array_vs_pos}
\end{figure}

The gain is not only measured inclusively, but also as a function of the position in the pads
by combining the beam telescope track position at the DUT $z$ coordinate with the signal on the LGAD detector (see Figures~\ref{fig:Sgain_vs_pos}, \ref{fig:gain_array_vs_pos}).
The gain is derived  for each DUT position from the charge distribution with the same procedure used for the inclusive measurement.
The circular structure in  the central part of the S1M-1 sensor has a slightly  lower gain than the external part of the DUT.
This shape corresponds  to  the opening in the metal layer used for laser testing, where a small potential drop is expected.

\subsubsection{Rise time and noise jitter}

\begin{figure}[htbp]
  \centering
  \subfloat[]{
   \includegraphics[width=0.45\textwidth]{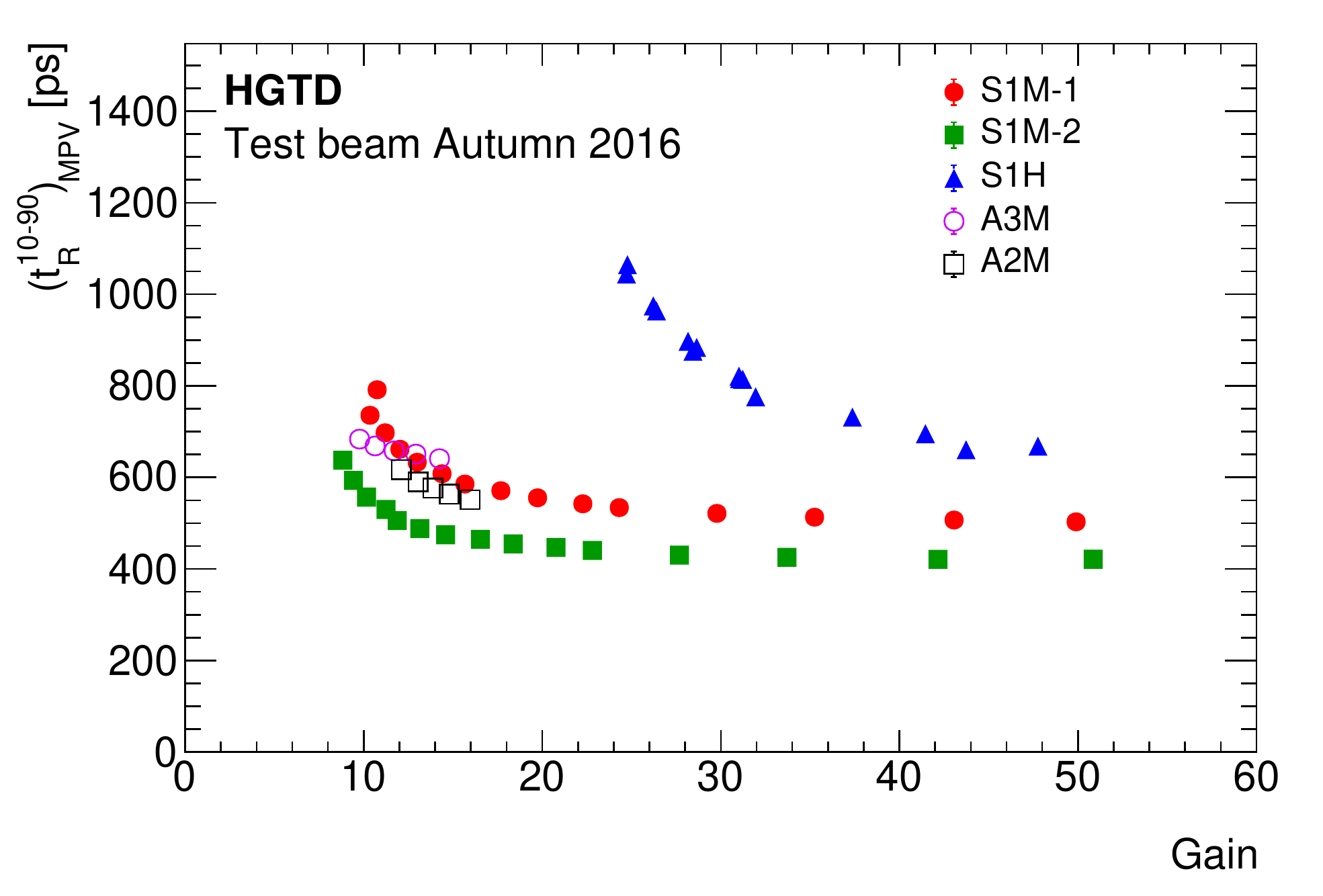}
    \label{fig:RiseTime}
  }
  \hfill
  \subfloat[]{
    \includegraphics[width=0.45\textwidth]{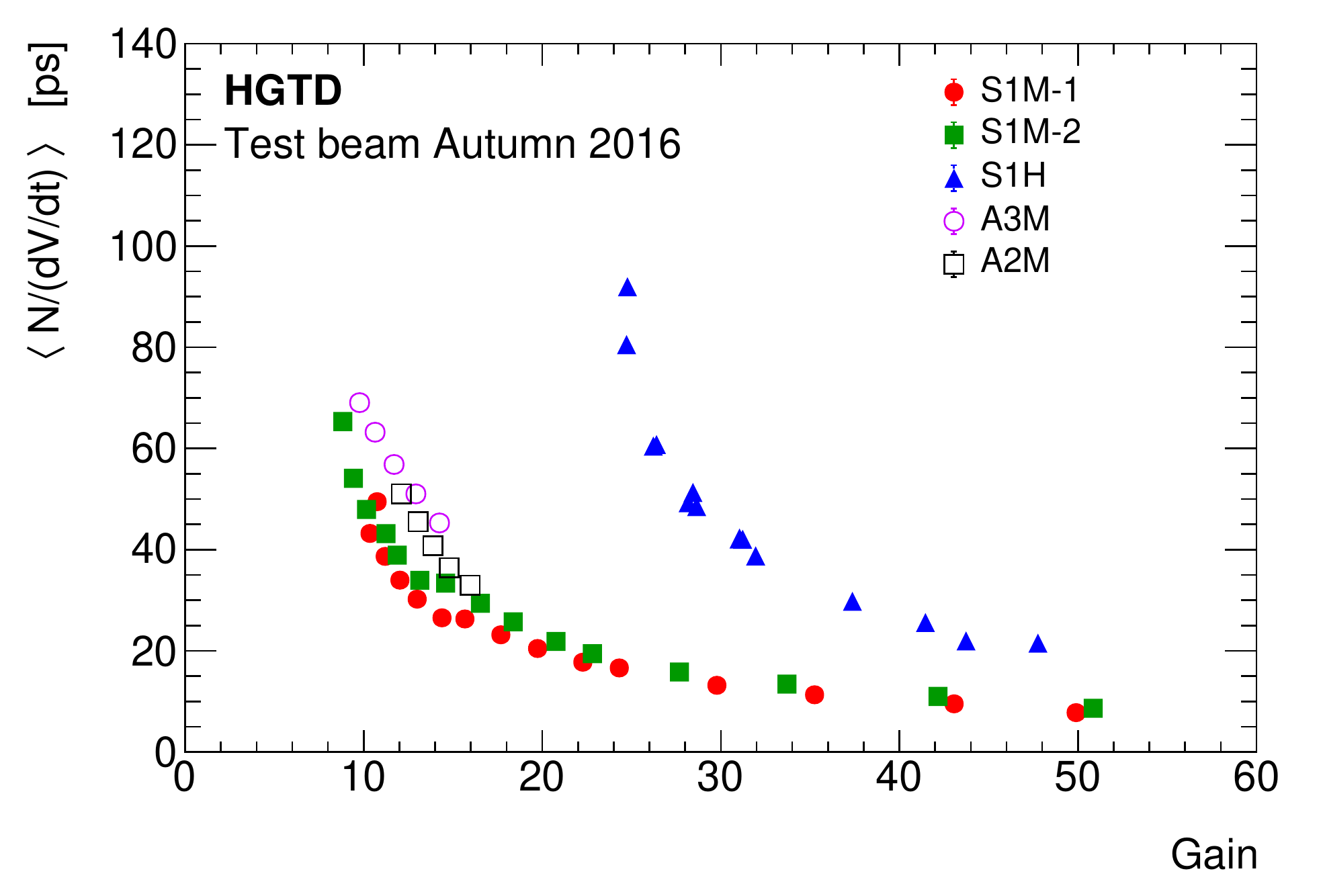}
    \label{fig:Jitter}
  }
  \caption{Rise time \protect\subref{fig:RiseTime} and jitter \protect\subref{fig:Jitter} as a function of the gain for various sensors.  Statistical uncertainties are negligible and smaller than the marker size.}
  \label{fig:RiseTimeJitter}
\end{figure}

The rise time, computed as the elapsed time from 10\% to 90\% of the pulse amplitude, is a critical parameter of a timing device. For a given signal-to-noise ratio,
the faster the rise time is, the better the time resolution is.
The measured pulse rise times as a function of the gain are shown in Figure~\ref{fig:RiseTime} for single pads and arrays.
The rise time decreases when the bias voltage is increased, due to the faster drift mobility.
Sensors with higher doping have larger rise time for a given gain due to the lower operating bias voltage. For the sensors with medium doping, a larger rise time is measured for S1M-1 compared to S1M-2.
The fastest rise time measured with S1M-2 at large gain is 420 ps.

%

One of the contributions to the LGAD time resolution is the electronic jitter defined as $N/(dV/dt)$ where $N$ is the electronic noise and $dV/dt$ is the derivative of the pulse when the signal crosses the threshold. The measured jitter values as a function of the gain are shown in Figure~\ref{fig:Jitter}. As expected, the jitter decreases with increasing gain.
The high doping sensors exhibit a larger jitter for the same gain value mostly correlated to the larger rise time.
The smallest jitter (8\,ps) is obtained for single pads with medium doping at the largest gain.
At a gain of 14 (the largest value obtainable for A3M), the jitter decreases from 45 ps at large capacitance value (22\,pF for A3M) to 26 ps for the smaller ones (3.9\,pF for S1).

It should be noted that the jitter depends on the time reconstruction method.
For the ZCD method, $N$\footnote{For the ZCD pulse, the noise is given by $N\cdot\sqrt{1+ f_{ZCD}^{2}-2 f_{ZCD}c(d_{ZCD})}$ where $c$ is the noise autocorrelation
 function. For all sensors, $|c(d_{ZCD})|$ is smaller than 0.1 except for S1M-1
 which has  $c(d_{ZCD})\sim0.2$.} and $dV/dt$ should be computed for the ZCD
pulse, i.e.~the difference of the original pulse and the delayed and attenuated
 copy. The jitter computed for the ZCD method is found to agree within a few
percent with the jitter computed for the CFD method with few exceptions.
 For S1M-2, the ZCD jitter is found to be about 20\% larger than the CFD
jitter at high gain and for A2M the ZCD jitter is found to be about 7\%
smaller than the CFD jitter.


\label{sec:properties}

\subsection{Efficiency}

\begin{figure}[htbp]
\centering
  \begin{tabular}{ll}
  \subfloat[]{
    \includegraphics[width=0.45\textwidth]{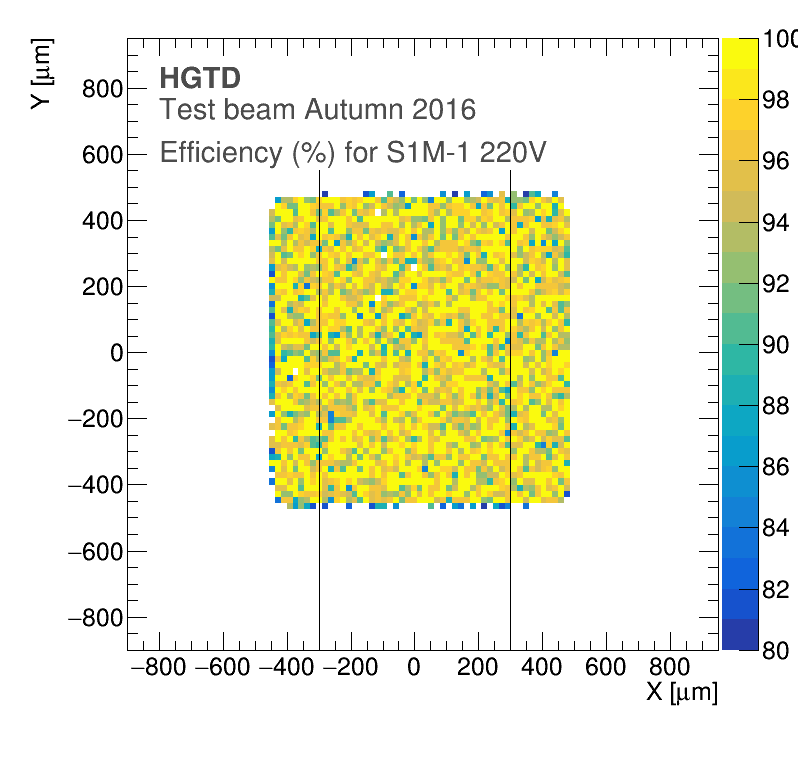}
    \label{fig:EfficLGA31}
  }
  \hfill
  \subfloat[]{
    \includegraphics[width=0.45\textwidth]{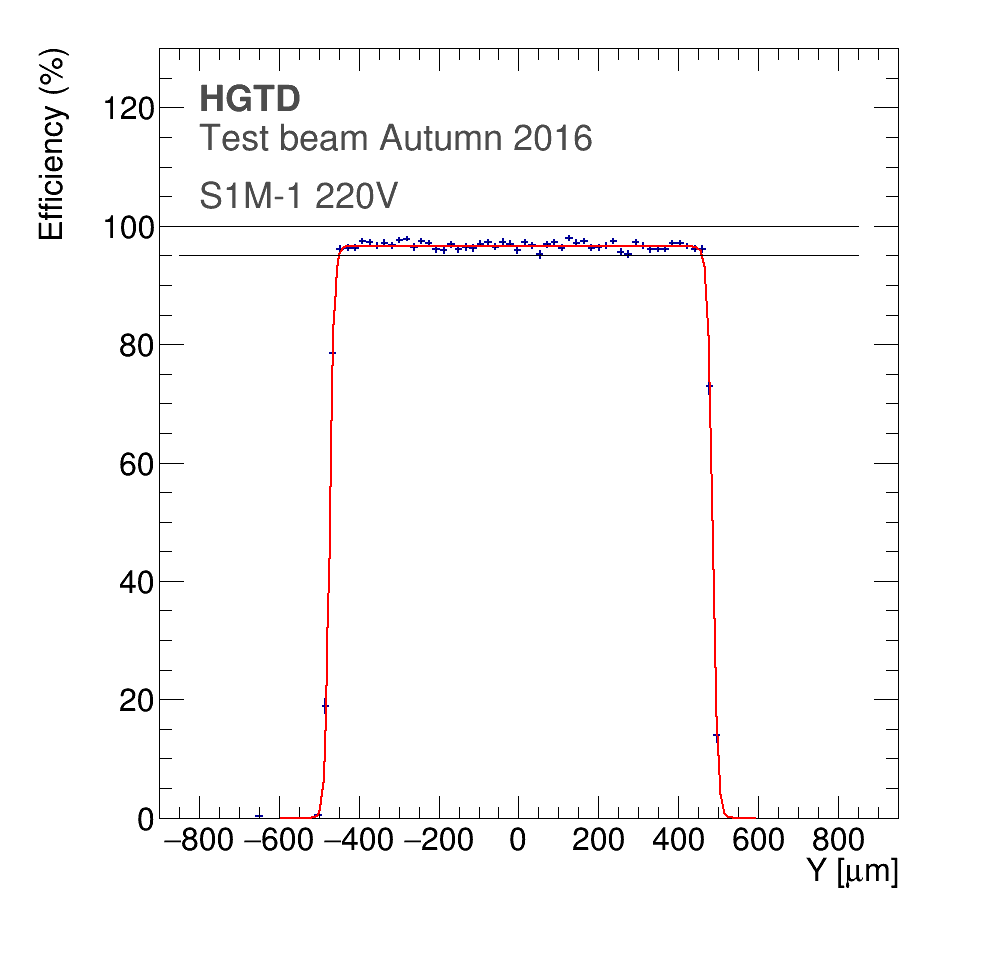}
    \label{fig:EfficVsYLGA31}
  } \\
  \end{tabular}
  \begin{tabular}{ll}
  \subfloat[]{
    \includegraphics[width=0.45\textwidth]{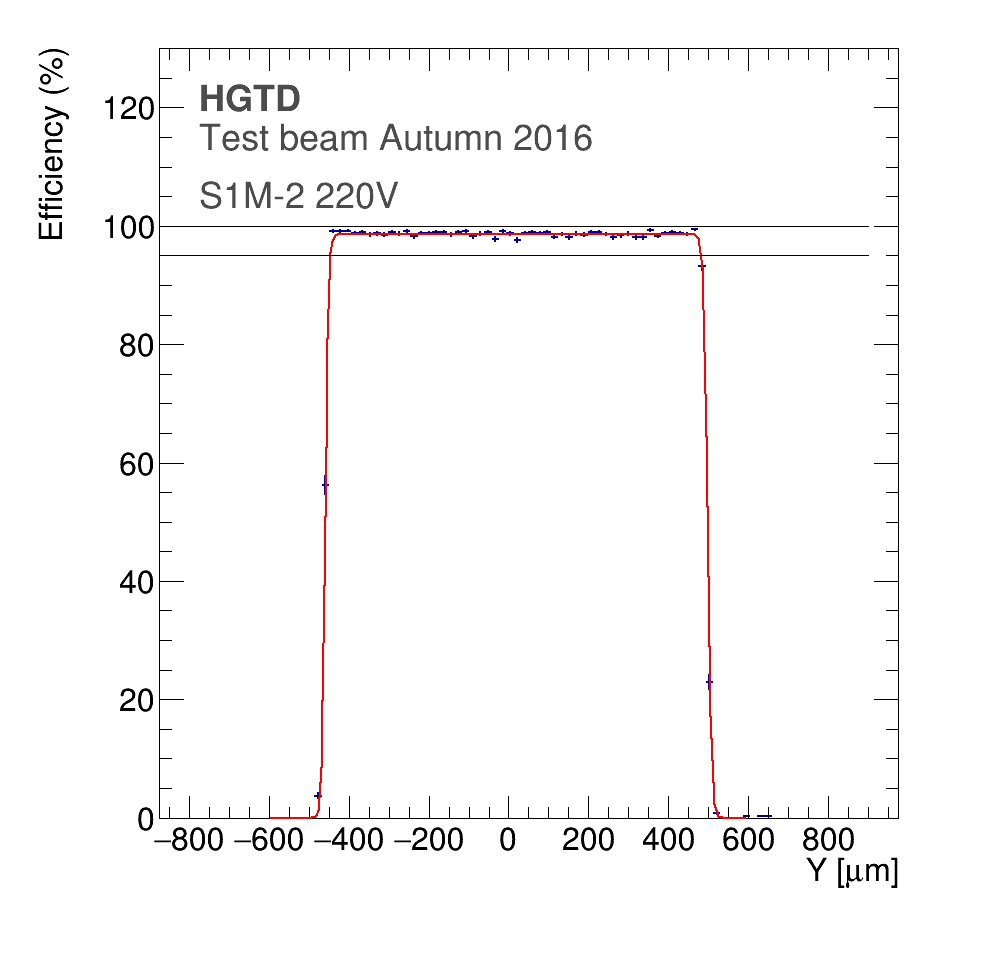}
    \label{fig:EfficVsYLGA33}
  }
  \hfill
  \subfloat[]{
    \includegraphics[width=0.45\textwidth]{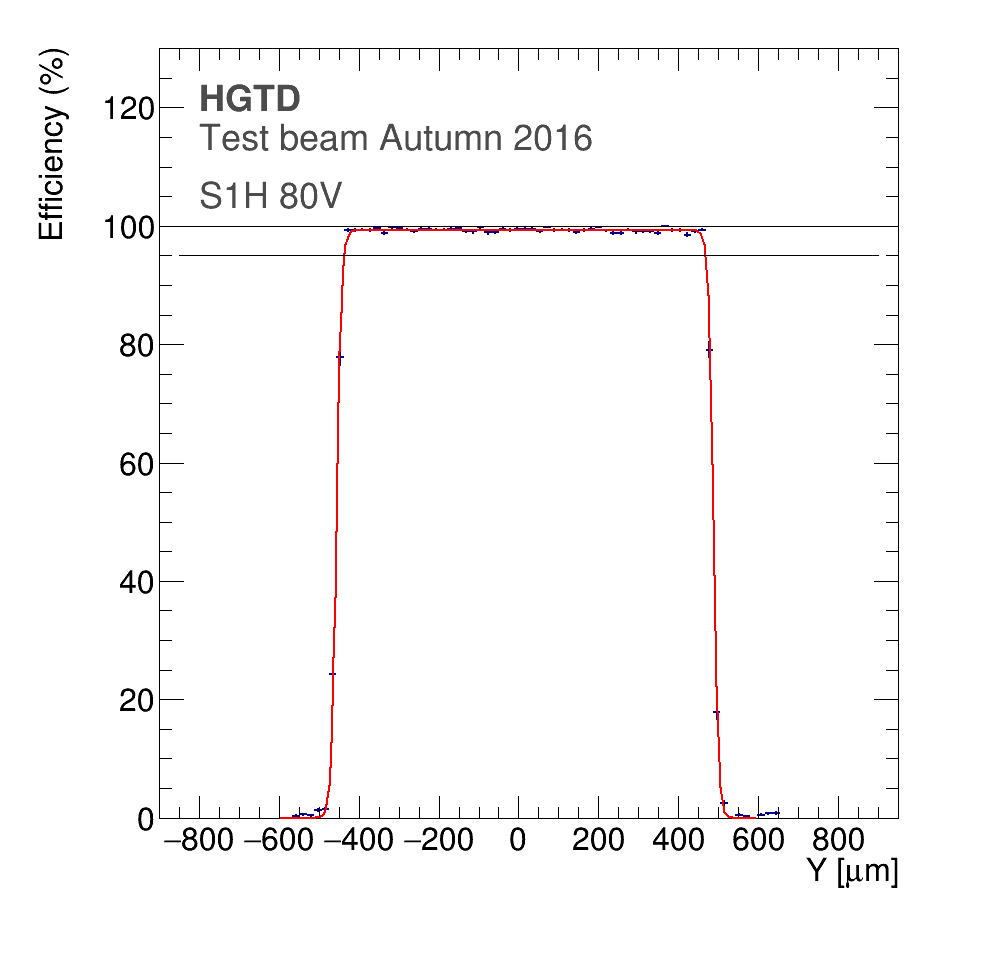}
    \label{fig:EfficVsYLGA34}
  } \\ 
  \end{tabular}
  \caption{Efficiency in percent for the single-pad sensor S1M-1, as a function of the position on the pad~\protect\subref{fig:EfficLGA31}. The bin size is $(18.5~\SI{}{\micro\meter})^2$.
Projections on the $y$-axis of the efficiency in the central region
(defined by the lines in~\protect\subref{fig:EfficLGA31}) in sensors
S1M-1~\protect\subref{fig:EfficVsYLGA31},
S1M-2~\protect\subref{fig:EfficVsYLGA33} and
S1H~\protect\subref{fig:EfficVsYLGA34}. The projections are fitted
with sigmoid functions (red line). 
}
  \label{fig:Seffic_vs_pos}
\end{figure}

\begin{figure}[htbp]
\centering
  \begin{tabular}{ll}
  \subfloat[]{
    \includegraphics[width=0.45\textwidth]{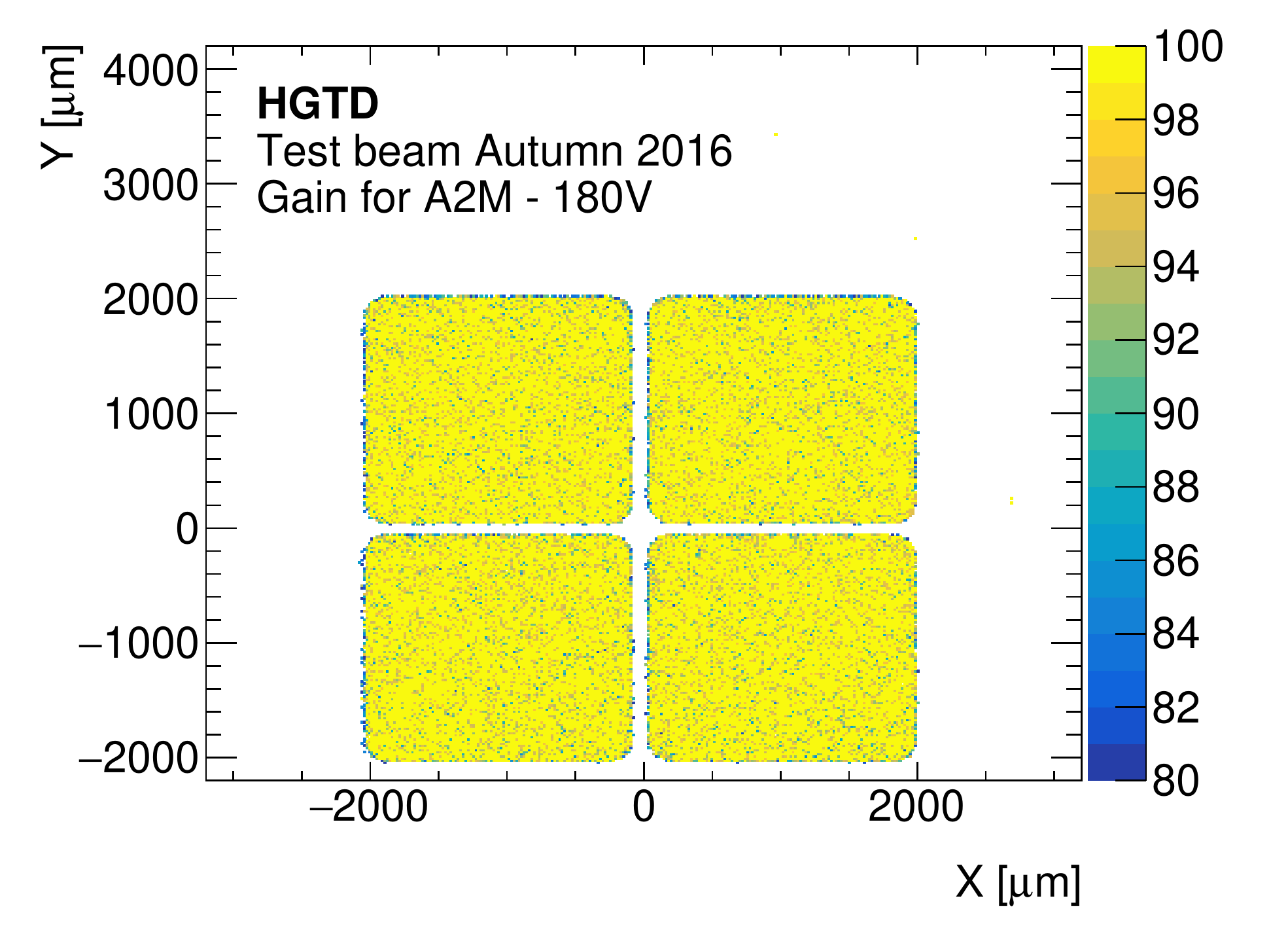}
    \label{fig:EfficA2M}
  }
  \hfill
  \subfloat[]{
    \includegraphics[width=0.45\textwidth]{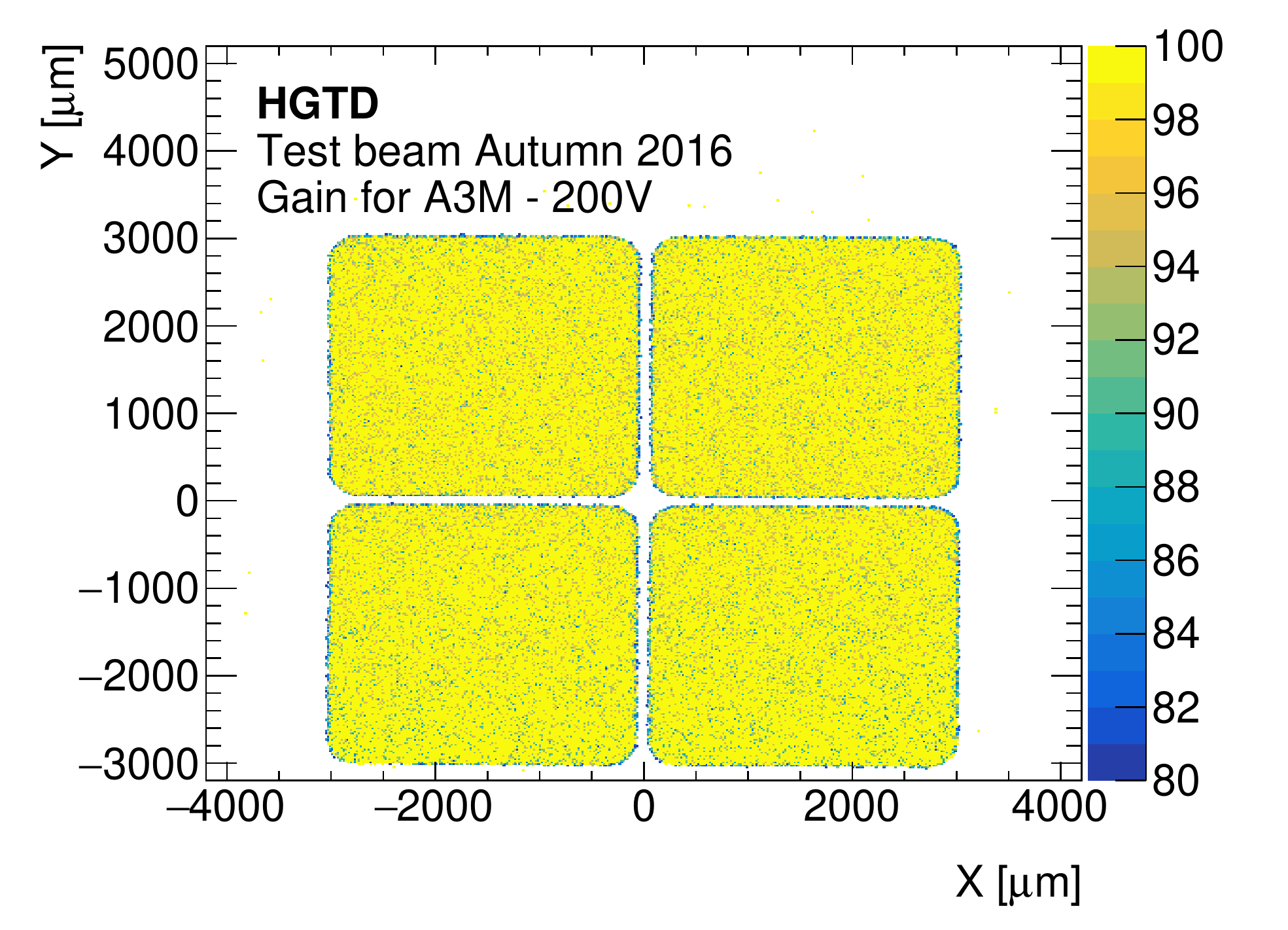}
    \label{fig:EfficA3M}
  }
  \end{tabular}
\caption{Efficiency in percent for the A2M \protect\subref{fig:EfficA2M}, A3M \protect\subref{fig:EfficA3M} sensor arrays as a function of the position on the pad. The bin size is $(18.5~\SI{}{\micro\meter})^2$.}
\label{fig:EfficArrays}
\end{figure}

\begin{figure}[htbp]
\centering
  \begin{tabular}{ll}
  \subfloat[]{
    \includegraphics[width=0.45\textwidth]{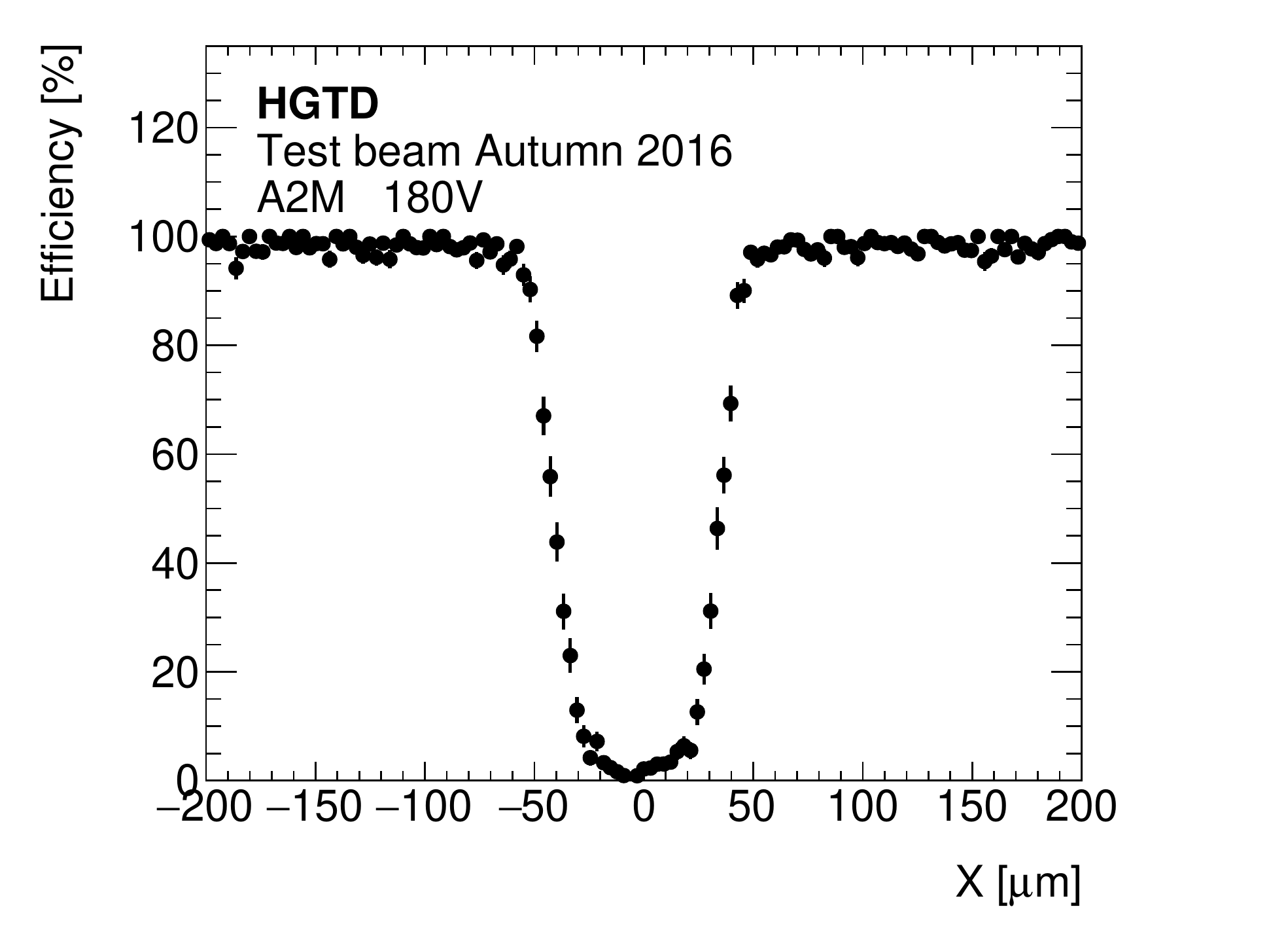}
    \label{fig:Effic34A2M}
  }
  \hfill
  \subfloat[]{
    \includegraphics[width=0.45\textwidth]{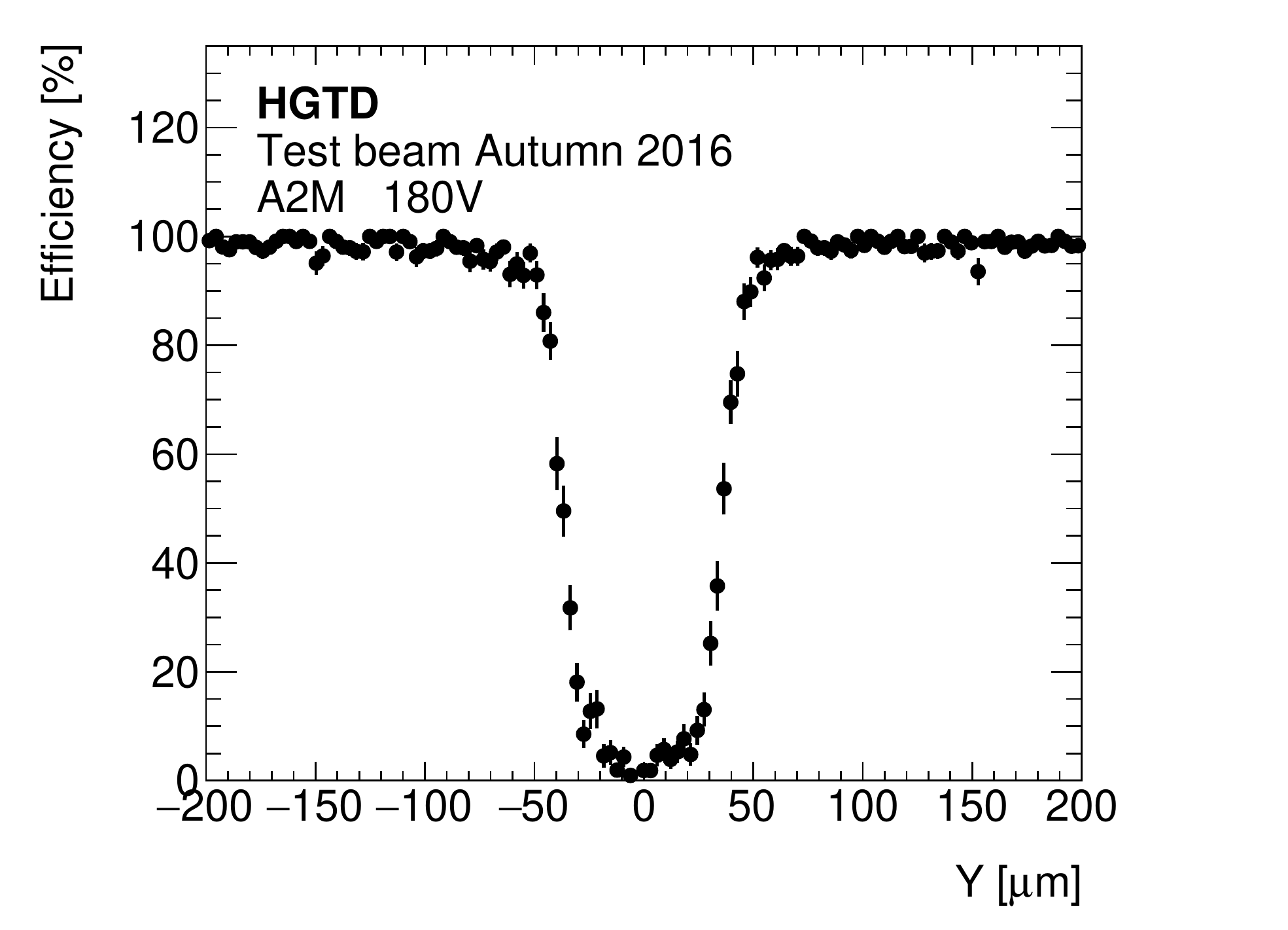}
    \label{fig:Effic13A2M}
  }\\
  \end{tabular}
  \caption{Efficiency in percent along the interpad gap of the A2M sensor in the $x$ direction \protect\subref{fig:Effic34A2M} and $y$ direction \protect\subref{fig:Effic13A2M}.} 
  \label{fig:eff_proj_A2M}
\end{figure}

The efficiency at a given position in the pad is defined as the number of hits that induce a sensor 
response (with amplitude above threshold) divided by the total number of reconstructed tracks crossing the DUT at that position. The amplitude threshold, given in Table~\ref{tab:effic_singlepad}, is different for each sensor and lies around the minimum between the noise and the signal peaks. The bin size is chosen in order to have a statistical uncertainty in each bin of about 2\%. The measured 2D distribution for one single-pad sensor as well as the projections in the $y$ direction are shown in Figure~\ref{fig:Seffic_vs_pos}. These projections are fitted with sigmoid functions. Figure~\ref{fig:EfficArrays} shows the corresponding 2D efficiency distribution for the arrays, while in Figure~\ref{fig:eff_proj_A2M} an example of projections along the $x$ and $y$ axes can be seen.
In the central region of the sensors, the mean values of the plateau efficiency and its dispersions (defined as 
the RMS of the efficiency distribution on the plateau) are summarised
in Table~\ref{tab:effic_singlepad}. The same table shows also the size
of regions where the efficiency is larger than 99.9\% or 50\% of the
plateau efficiency. For the arrays, the width of the inter-pad region
for each sensor, defined as the region with efficiency below 50\%, is estimated
to be (76$\pm$5)\,\SI{}{\micro\meter}, slightly larger than the expected
width of 63\,\SI{}{\micro\meter}. The width estimate is performed by
parametrising the edges with Gaussian functions and determining the
50\% efficiency point based on the fit parameters and their uncertainties.



\begin{table}[htbp]
  \centering
  \caption{Mean efficiency and its dispersion on the plateau for each sensor or pad. The threshold used to compute these efficiencies is also given. The last two columns contain the size of the regions where the efficiency is larger than 99.9\% or 50\% of the efficiency given in column 3, respectively.}
  \begin{tabular}{|c|c|c|c|c|c|}
\hline
 Sensor & Threshold amplitude  & Efficiency  & Dispersion &  \multicolumn{2}{c|}{Size of the plateau (\SI{}{\micro\meter})}     \\
 &for the signal & on the plateau & & at 99.9\% & at 50\% \\
\hline
 S1M-1 & 60~mV & $(96.7\pm0.1)$\%  & 0.7\%  & 876& 960  \\
 S1M-2 & 40~mV & $(98.6\pm0.1)$\% & 0.4\%  & 898 & 958 \\
 S1H   & 30~mV & $(99.3\pm0.1)$\% & 0.2\%  & 859 & 945 \\ 
 A2M & 40~mV & $(96.0\pm0.1)$\% & 1.1\%  & 1920 & 1995 \\
 A3M & 25~mV & $(97.0\pm0.1)$\% & 1.0\%  & 2930 & 3006 \\ 
\hline
  \end{tabular}
  \label{tab:effic_singlepad}
\end{table}

\subsection{Crosstalk hit probability}
Events with signal above a certain threshold on more than one pad were studied independently to estimate detector induced crosstalk hit probability in arrays. The corresponding rate was defined as the ratio of coincidences between two neighbouring pads over the sum of events in each individual pad, reduced by the number of coincidences. The
final number is estimated according to Eq.~\ref{eq:XTalR}:

\begin{equation}  \label{eq:XTalR}
 R_{XTalk} = n_{ij}/(n_{i}+n_{j}-n_{ij})
\end{equation}

where the indices $i$ and $j$ denote the corresponding pads. In this section a hit is defined as any event with a signal amplitude
over 30\,mV in any of the examined pads. The crosstalk ratio is estimated to be $0.3\%$ and $0.6\%$ for
the A2M and A3M DUTs, respectively. The common threshold was set at 30\,mV threshold to be well within the gap between the noise and the signal amplitude peaks in each pad of both DUTs.

\begin{figure}[htbp]
  \begin{tabular}{ll}
    \subfloat[]{
    \includegraphics[width=0.45\textwidth]{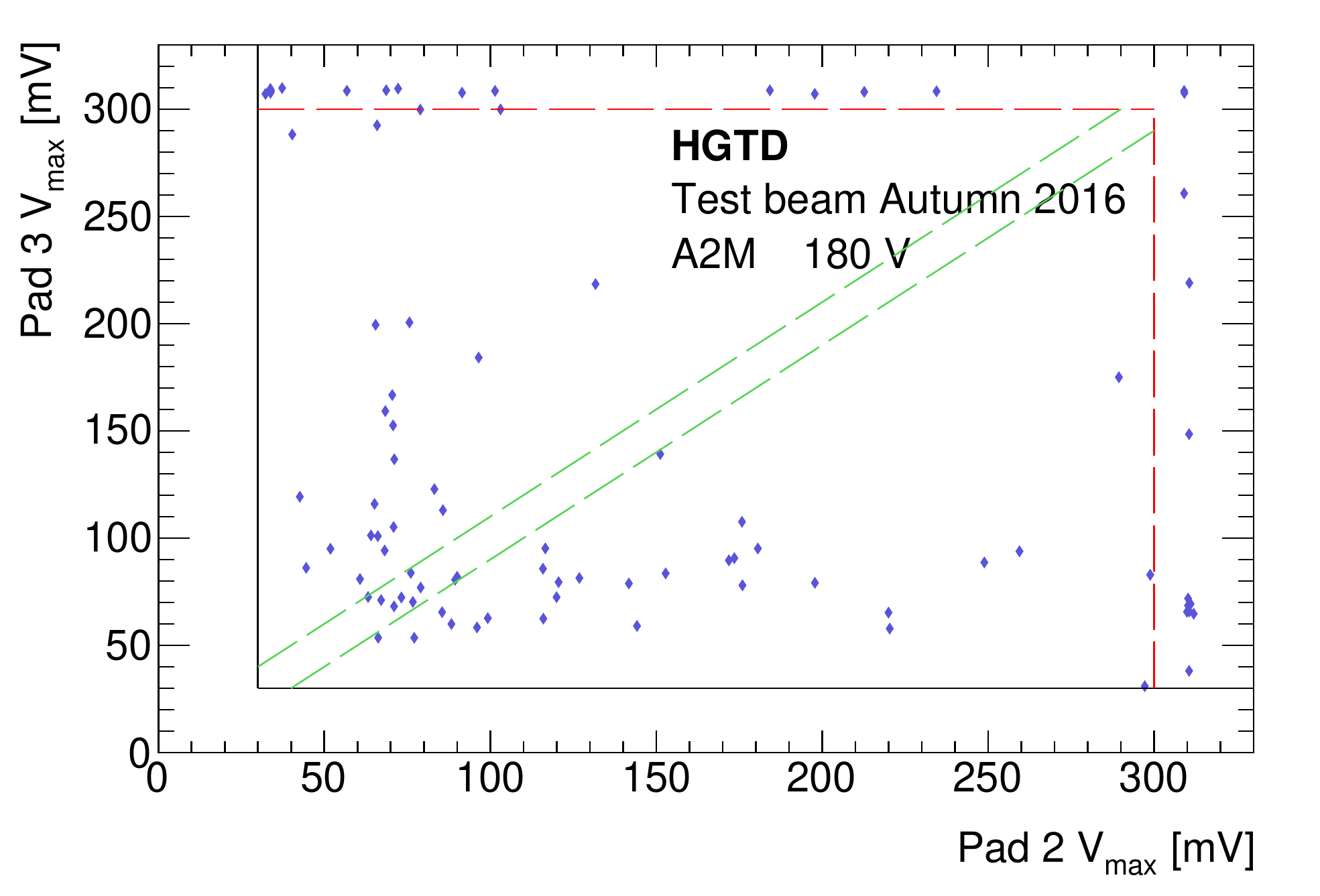}
    \label{fig:XtalkHits}
    }
    \hfill
    \subfloat[]{
    \includegraphics[width=0.45\textwidth]{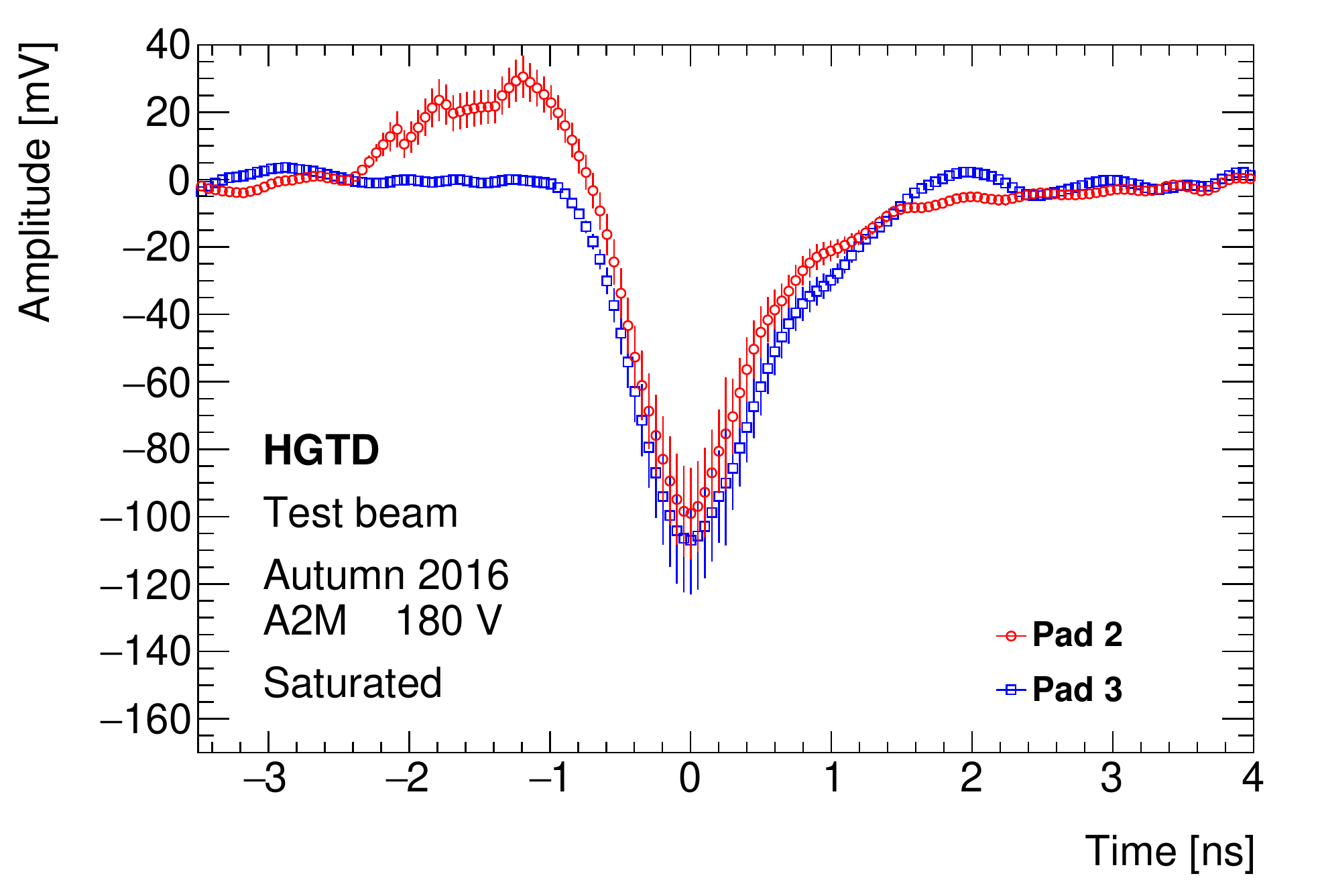}
    \label{fig:XtalkSat}
    }\\
  \end{tabular}
  \begin{tabular}{ll}
    \subfloat[]{
    \includegraphics[width=0.45\textwidth]{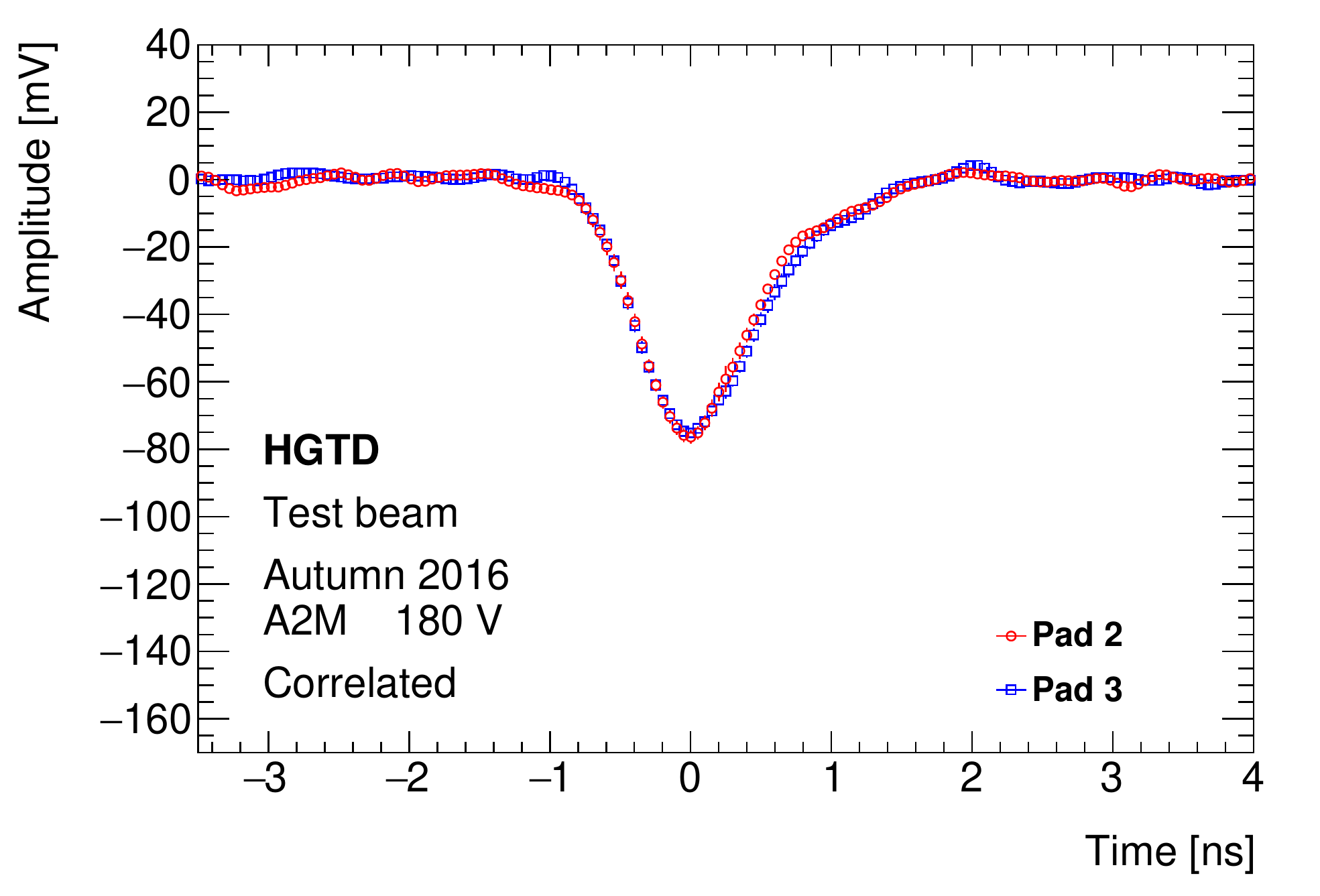}
    \label{fig:XtalkCor}
    }
    \hfill
    \subfloat[]{
    \includegraphics[width=0.45\textwidth]{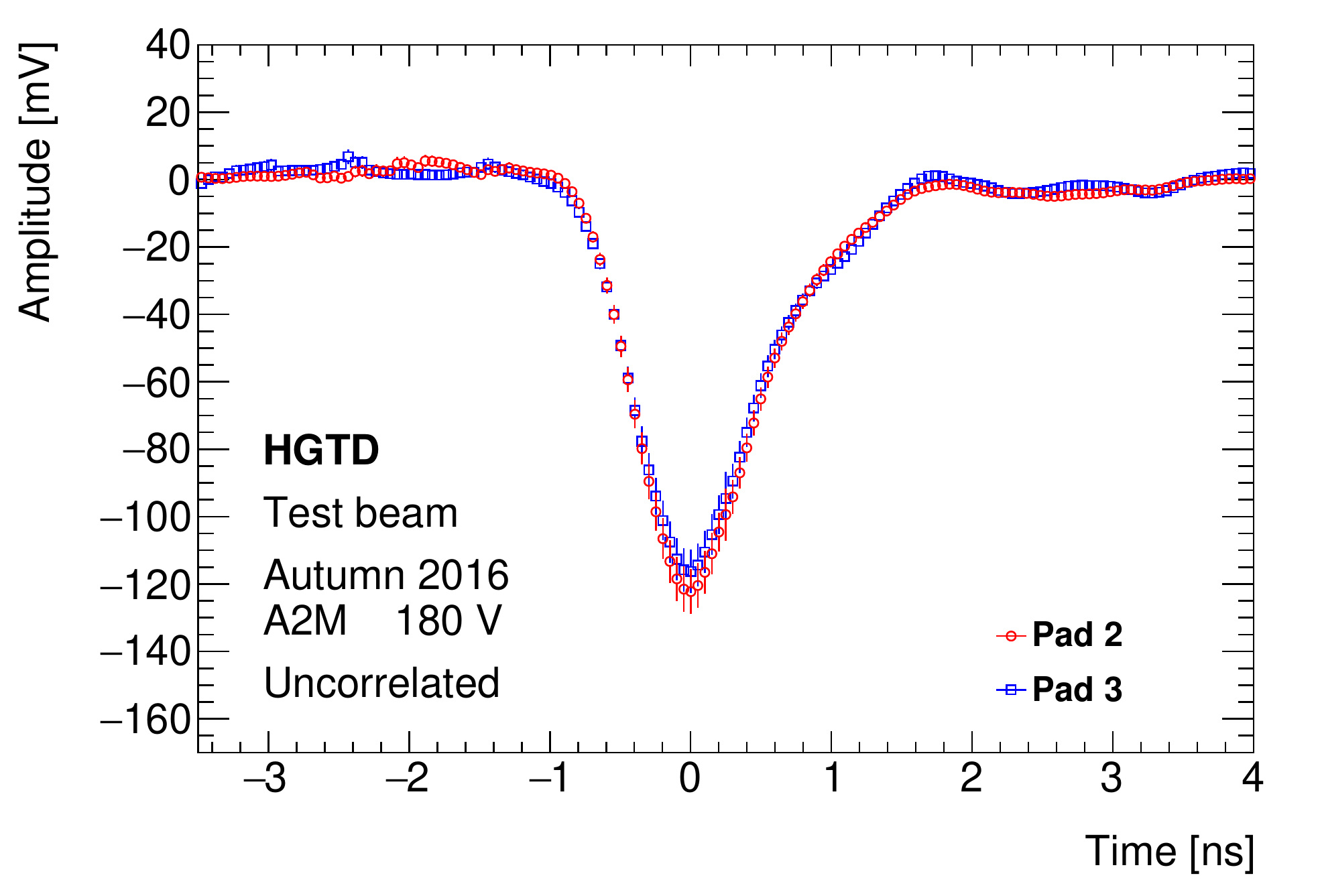}
    \label{fig:XtalkUnc}
    }\\
  \end{tabular}
  \caption{Correlation between the signal amplitudes $V_{\text{max}}$ of neighbouring channels with signals within 2\,ns from each other~\protect\subref{fig:XtalkHits} for the A2M array. The black lines indicate the threshold to define a hit (signals with smaller amplitude are not shown), the red lines the minimal amplitude on one pad for saturated events. Correlated hits lie within the green lines, uncorrelated ones lie outside of them, but below the red lines. Average waveforms in each of the pads for saturated events with amplitude above 300 mV in the other pad and below 300 mV in the considered one~\protect\subref{fig:XtalkSat}, for correlated crosstalk~\protect\subref{fig:XtalkCor} and uncorrelated crosstalk~\protect\subref{fig:XtalkUnc}.}
 \end{figure}

Considering the amplitudes and time differences of the signals, three regions can be identified originating from different sources. All events are required to be in-time, i.e.~their maximum amplitude must lie within $\delta t<2$\,ns. Double hits can show saturation in either or both pads (``saturation events''), similar amplitudes in both pads (``correlated crosstalk'') due to a particle hitting the interpad region, or different amplitudes (``uncorrelated crosstalk'') .  Figure~\ref{fig:XtalkHits} shows the amplitudes $V_{\textrm{max}}$ in the two hit pads with $V_{\textrm{max}}$ > 30 mV and $\delta t<2$\,ns.

\begin{description}
\item[Saturation events] This region refers to events with amplitude close to the highest possible value that can be recorded by the oscilloscope at the chosen settings in one channel, accompanied by a secondary hit in a neighbouring channel with lower amplitude. The channel where the highest amplitude is reached, referred to as the primary channel, results in an increased charge injection within the bulk that distorts the electrical field. This field distortion may propagate to the closest neighbour inducing a secondary signal that will typically be of smaller amplitude but within the same time window as the primary event.
Double hit events are considered to belong to this category if the signals have an amplitude of at least 300\,mV in at least one of the channels while being within a 2\,ns time window of each other.
They represent 0.1\% and 0.2\% of the total number of hits for the A2M and A3M DUTs, respectively, and in the time of arrival vs.~amplitude plane, they are mainly positioned around the $\delta t\simeq 0$\,ns line, while amplitudes of the secondary hit are spread from the lower to the higher extremities of the distribution. The average signal shape in the non-saturated pad of A2M in this kind of events is shown in Figure~\ref{fig:XtalkSat}.
\item[Correlated crosstalk] In the case of an event occurring in the space between pads, a signal is induced in both pads with a ratio of amplitudes directly related to the distance of the hit to each pad border. This type of events is easily identifiable in the special case where the original hit occurred close to the middle of the inter-pad region. Recorded signals will be on time with one another and of similar amplitudes, thus mainly concentrated on the diagonal of the amplitude correlation plot between different channels. As expected, these events represent a very low fraction of the total recorded hits, corresponding to 0.05\% and 0.1\%  for the A2M and A3M DUTs, respectively. For A2M, their amplitude lies mostly around 75 mV as shown by the average pulse in Figure~\ref{fig:XtalkCor}.
\item[Uncorrelated crosstalk] The case of inter-pad coincidences where signals do not present any clear correlation is included in this category. The timing requirement of $\delta t<2$\,ns is still applied, while to exclude correlated coincidences, the amplitudes of the two signals are required to differ by at least 10\,mV. Given the relatively large induced signal amplitudes, these events are not expected to originate from any non-gain region of the DUT, such as the periphery or the space between neighbouring pads.  They represent approximately half of the observed crosstalk in both the A2M and A3M DUTs, corresponding to a probability of 0.15\%  and 0.3\%, respectively. Figure~\ref{fig:XtalkUnc} shows their average pulse with an amplitude similar to the saturated events, but with a much smaller spread.
\end{description}

A summary of all calculated crosstalk hit probabilities for the A2M and A3M DUTs is presented in Table \ref{tab:XTalkSummary}. The twice as high level of crosstalk in A3M than in A2M is assumed to originate from the different read-out boards.

\begin{table}
\begin{center}
\begin{tabular}{|c|c|c|}
\hline
 \multicolumn{3}{|c|}{Crosstalk hit probability}\\
 \hline
  Category & A2M & A3M\\
 \hline
  Saturated & 0.10\% & 0.2\% \\
  Correlated & 0.05\% & 0.1\% \\
  Uncorrelated & 0.15\% & 0.3\% \\
\hline
  Total & 0.30\% & 0.6\% \\
\hline
\end{tabular}
\end{center}
\caption{Summary table of crosstalk hit probability for the A2M and A3M DUTs}
\label{tab:XTalkSummary}
\end{table}

\subsection{Time resolution}
\subsubsection{Optimization}
\label{sec:opti}

\begin{figure}[htbp]
  \centering
  \subfloat[]{
    \includegraphics[width=0.45\textwidth]{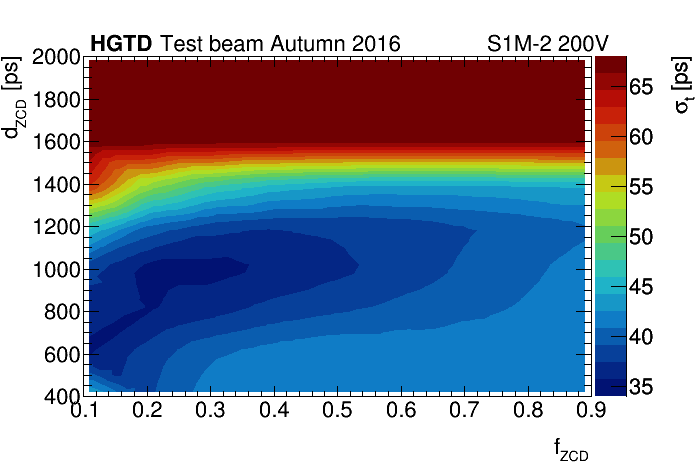}
    \label{fig:Opti2D_S1M-2}
  }
  \hfill
  \subfloat[]{
    \includegraphics[width=0.45\textwidth]{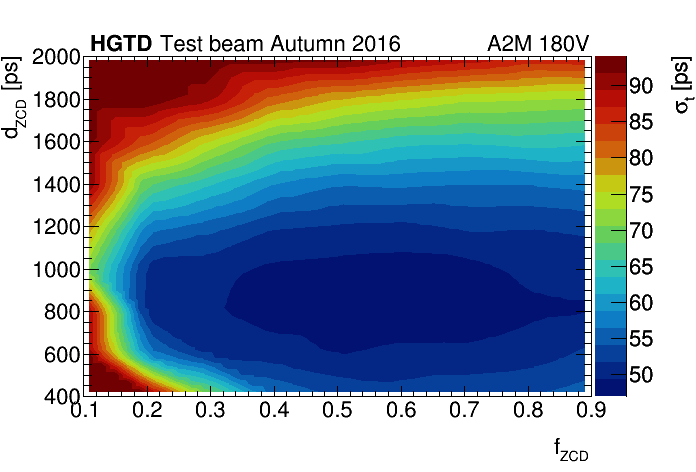}
    \label{fig:Opti2D_A2M}
  }
  \caption{Time resolution as a function of the constant fraction and delay parameters for S1M-2 at 200~V \protect\subref{fig:Opti2D_S1M-2} and A2M at 180~V \protect\subref{fig:Opti2D_A2M}.}
  \label{fig:Opti2D}
\end{figure}

\begin{figure}[htbp]
  \centering

  \subfloat[]{
    \includegraphics[width=0.45\textwidth]{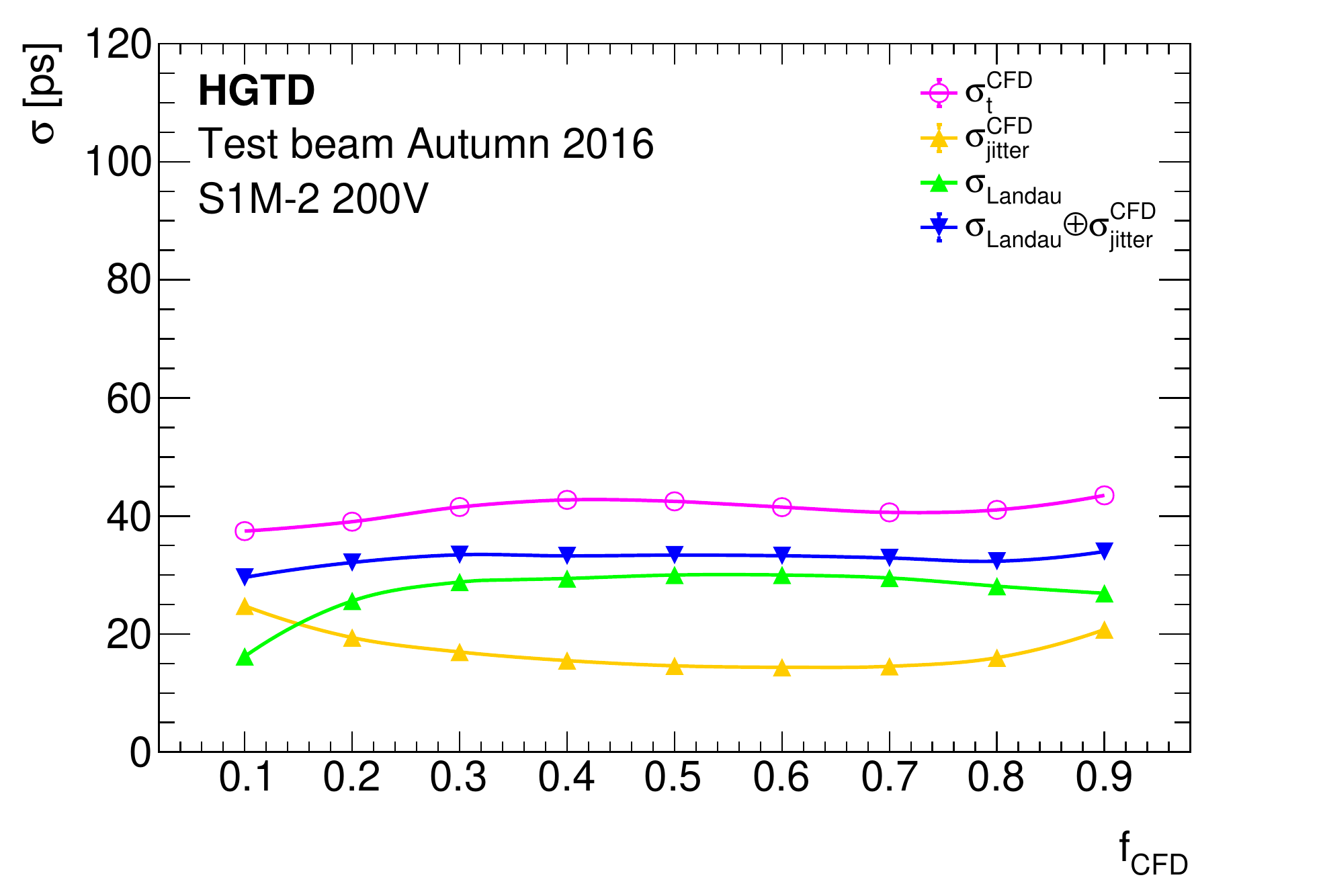}
    \label{fig:OptiCFD_S1M-2}
  }
  \hfill
  \subfloat[]{
    \includegraphics[width=0.45\textwidth]{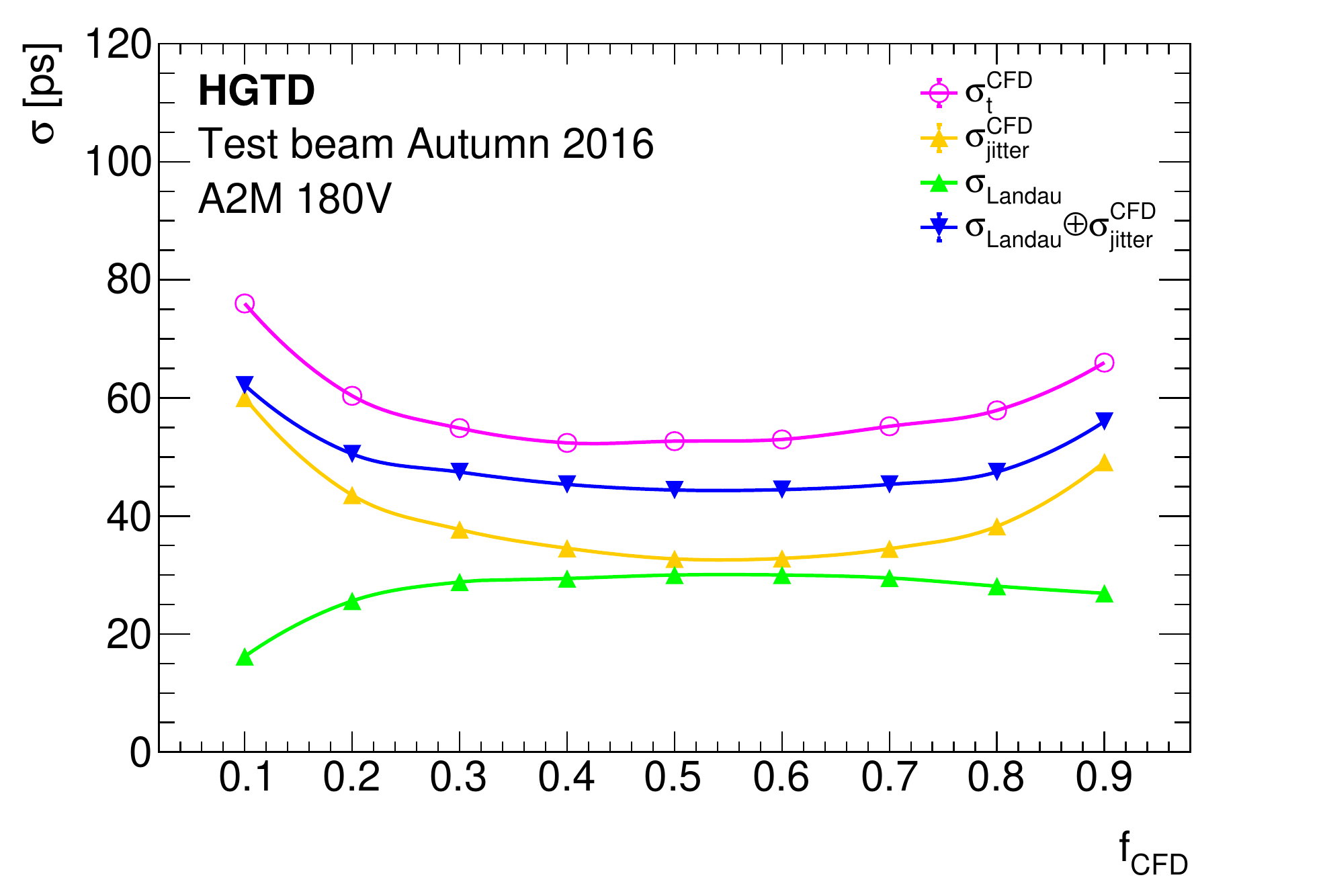}
    \label{fig:OptiCFD_A2M}
  }
  \caption{Time resolution  for the CFD method $\sigma_t^\textrm{CFD}$ as a function of the constant fraction parameters compared with the predictions for S1M-2 \protect\subref{fig:OptiCFD_S1M-2} and A2M \protect\subref{fig:OptiCFD_A2M}. The term $\sigma_{\mathrm{Landau}}$ is computed with \texttt{Weightfield2}~\cite{WF2,Cartiglia:2017mcj}, while $\sigma_{\mathrm{Jitter}}^\textrm{CFD}$ is estimated from data (see Section~\ref{sec:properties}).}
  \label{fig:Opti}
\end{figure}

The parameters of the CFD and ZCD algorithms have been chosen in order to optimise time resolution. The constant fraction was scanned in 10\% steps from 10\% to 90\% of the amplitude and the delay was scanned in 200\,ps steps from 400 to 2000\,ps. Figure~\ref{fig:Opti2D} shows two examples of two-dimensional resolution map for the ZCD method. For S1M-2 at 200\,V, the optimal value of $d_{ZCD}$ is close to 1\,ns, while a slightly lower value is found for A2M at 180~V. Similar results were obtained for other sensors and at different bias voltages. For simplicity, the same value $d_{ZCD}$=1\,ns is used everywhere, leading to slightly non-optimal time resolution at the level of 5\%. For S1M-2 at 200\,V, the optimal value of $f_{ZCD}$ is close to 0.2 while larger values, close to 0.5, are preferred for A2M. The conclusion is similar for the CFD method and the result agrees with the expectation obtained from the quadratic sum of the jitter and the Landau contributions, as shown in Figure~\ref{fig:Opti}. The time resolution for single pads at high bias voltage is dominated by the Landau term that prefers low values of the constant fraction~\cite{Cartiglia:2017mcj}. For arrays, the jitter contribution is dominant and this term prefers a larger constant fraction that maximizes $dV/dt$.
While for arrays the optimal value was found to be close to 0.5 at all bias voltages, for single pads and at high bias voltage (above 150~V for LGADs with medium doping and above 60~V for LGAD with high doping), the optimal value was found to be around 0.2. For lower bias voltage, slightly larger values  ($\sim$0.3) are preferred because the contribution of the jitter increases.
While the predictions qualitatively explain the preferred values of the constant fraction, some differences are observed between the measured time resolution and the prediction. The largest difference is observed for A3M with a measured time resolution 40\% higher than the prediction.

\subsubsection{Method comparison}

\begin{figure}[htbp]
  \centering
  \subfloat[]{
    \includegraphics[width=0.45\textwidth]{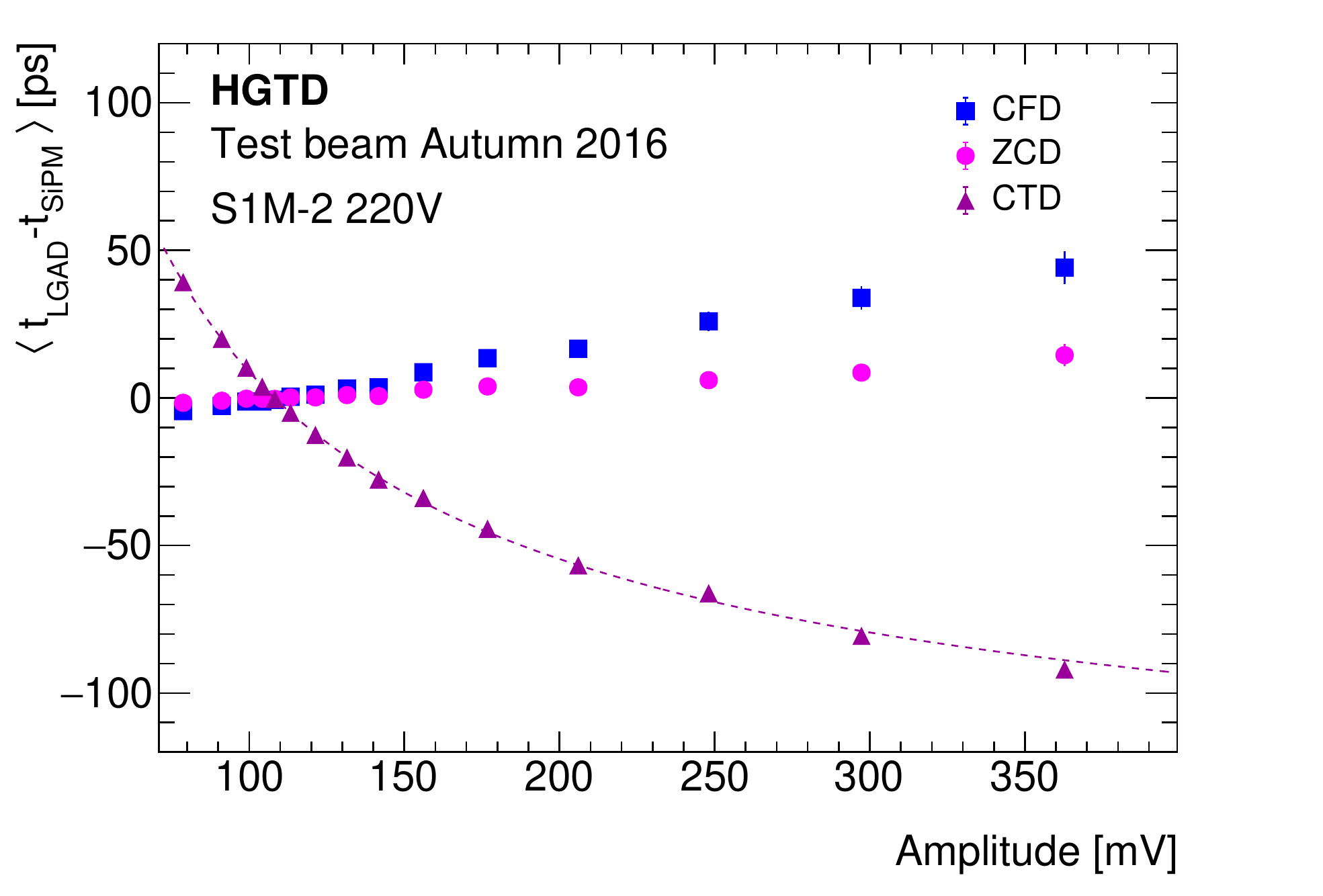}
    \label{fig:TvsA}
  }
  \hfill
  \centering
  \subfloat[]{
    \includegraphics[width=0.45\textwidth]{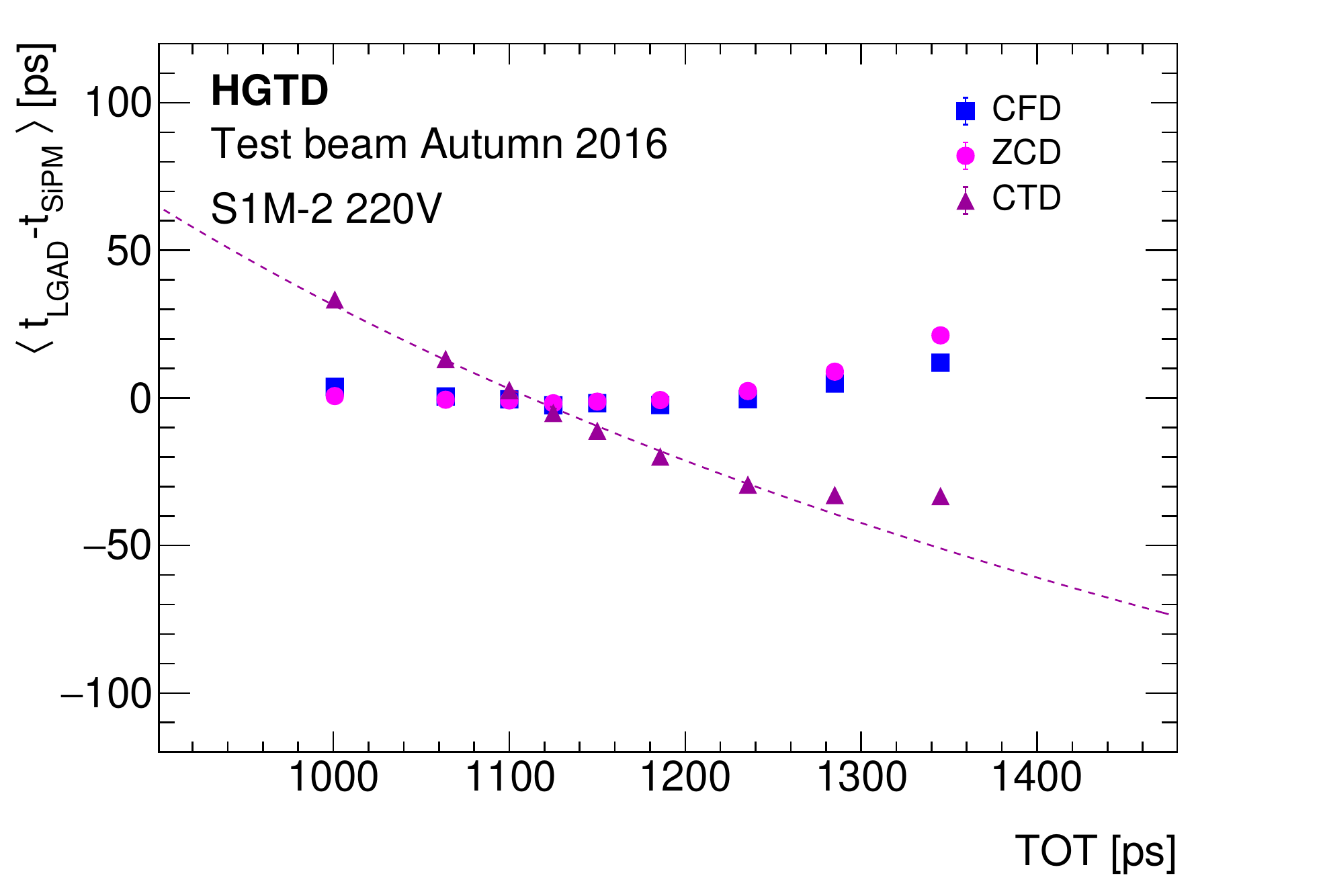}
    \label{fig:TvsTOT}
  }
  \caption{Average time difference between LGAD and SiPM versus pulse amplitude \protect\subref{fig:TvsA} and TOT \protect\subref{fig:TvsTOT} for various time reconstruction methods. The lines are fitted functions to the distributions for CTD.}
  \label{fig:Tdependance}
\end{figure}

The performance of the different time reconstruction algorithms has been compared for single pads and arrays with medium doping at high bias voltage.
The stability of the reconstructed time as a function of the amplitude has been investigated for the various methods.
Figure~\ref{fig:TvsA} shows the mean value of the reconstructed time for S1M-2 at 220\,V measured relatively to the SiPM time as a function of the reconstructed amplitude.
As expected, a larger time walk is observed for the CTD method with a threshold of 20\,mV, with a 50\,ps variation in amplitude between 100 and 200\,mV (most probable value is 100\,mV per MIP as shown in Figure~\ref{fig:Amplitude}).
While better behaving, the ZCD and CFD methods still exhibit an increasing time of arrival as a function of the amplitude.

The time resolution has been measured for the three methods. With the CFD method, the resolution is
(32.8$\pm$0.1)\,ps where the uncertainty is only statistical. A better resolution is measured with the ZCD
method, (29.3$\pm$0.1)\,ps, while a significantly larger resolution is measured for CTD (40$\pm$0.1)\,ps due to
the time walk effect. This bias can be minimized by correcting the CTD time as a function of the amplitude.
Using the fitted function shown in Figure~\ref{fig:TvsA}, the time resolution has been reduced by about 20\% leading
to a resolution of (29.9$\pm$0.1)\,ps. Similar corrections have been tested for the two
other methods, leading to improvements smaller than 5\%, as expected.
If the full pulse shape is not available, the TOT can be used as an estimate of the amplitude. Here a fixed threshold of 20~mV is used to compute this quantity. For S1M-2, and more generally for single
 pads close to breakdown voltage, distorted
pulses are observed leading to high TOT values uncorrelated with the amplitudes.
This is due to cases in which the holes in the amplification region start to show multiplication too. Distorted pulses close to the breakdown voltage are rejected using an upper cut on the TOT. This requirement rejects up to 5\% of the events at the highest bias voltage.
Figure~\ref{fig:TvsTOT} shows the mean value of the reconstructed time for S1M-2 as a function of the TOT.
A smaller dependence is observed because the correlation factor between the TOT and the amplitude is only 0.60.
The time resolution with TOT correction is (31.3$\pm$0.1)\,ps which is 5\% worse than the time resolution with the amplitude correction.
Similar analyses were performed for other sensors and at different bias voltages. The results are summarized in Table~\ref{tab:comparison}.
The best performance is obtained with the ZDC algorithm, chosen as the default algorithm in this paper, but difficult to implement in an ASIC. Currently the CTD method is used in the ASIC with offline correction of the timewalk with the TOT information. For single pad sensors at all bias voltages the time resolution measured with the CTD method with TOT correction is less than 7\% worse than with the ZCD method, in some cases even slightly better.


\begin{table}
 \caption{\label{tab:comparison} Time resolution for various sensors, various bias voltages and various time reconstruction methods. }
  \small
  \begin{center}
    \begin{tabular}{|c|c|c|c|c|c|c|}
      \hline
             &  &   \multicolumn{5}{c|}{Time resolution [ps]}\\
             &  & & &  \multicolumn{3}{c|}{CTD}\\
    Sensor & voltage & CFD & ZCD & no & amplitude & TOT\\
             & (V)  &  & & correction & correction & correction\\
             \hline

        & 200  &  38.6 $\pm$ 0.1 &  36.9 $\pm$ 0.1 &  49.4 $\pm$ 0.1 &  34.8 $\pm$ 0.1  &  39.3 $\pm$ 0.1 \\
  S1M-1 & 220  &  34.2 $\pm$ 0.1 &  33.1 $\pm$ 0.1 &  38.3 $\pm$ 0.1 &  29.4 $\pm$ 0.1  & 33.0 $\pm$ 0.1  \\
        & 240  &  27.1 $\pm$ 0.1 &  27.4 $\pm$ 0.1 &  30.8 $\pm$ 0.1 &  25.4 $\pm$ 0.1  & 27.9 $\pm$ 0.1  \\
             \hline
        & 200  &  38.3 $\pm$ 0.1 &  34.7 $\pm$ 0.1 &  55.1 $\pm$ 0.2 &  36.8 $\pm$ 0.2  & 37.0 $\pm$ 0.2  \\
  S1M-2 & 220  &  32.8 $\pm$ 0.1 &  29.3 $\pm$ 0.1 &  40.0 $\pm$ 0.1 &  29.9 $\pm$ 0.1  & 31.3 $\pm$ 0.1  \\
        & 240  &  27.9 $\pm$ 0.1 &  27.8 $\pm$ 0.1 &  31.6 $\pm$ 0.1 &  25.5 $\pm$ 0.1  &  27.5 $\pm$ 0.1  \\
             \hline
 \multirow{2}{*}{A2M}  & 140  &  73.9 $\pm$ 0.5 &  63.4 $\pm$ 0.5 &  102.0 $\pm$ 0.7 &  69.5 $\pm$ 0.7  & 64.3 $\pm$ 0.7  \\
       & 180  &  51.1 $\pm$ 0.3 &  47.0 $\pm$ 0.3 &  74.1 $\pm$ 0.5 &  51.8 $\pm$ 0.5  &   54.4 $\pm$ 0.5 \\

   \hline

      \hline
    \end{tabular}
  \end{center}
\end{table}

\subsubsection{Gain dependence}

\begin{figure}[htbp]
  \centering
  \subfloat[]{
    \includegraphics[width=0.45\textwidth]{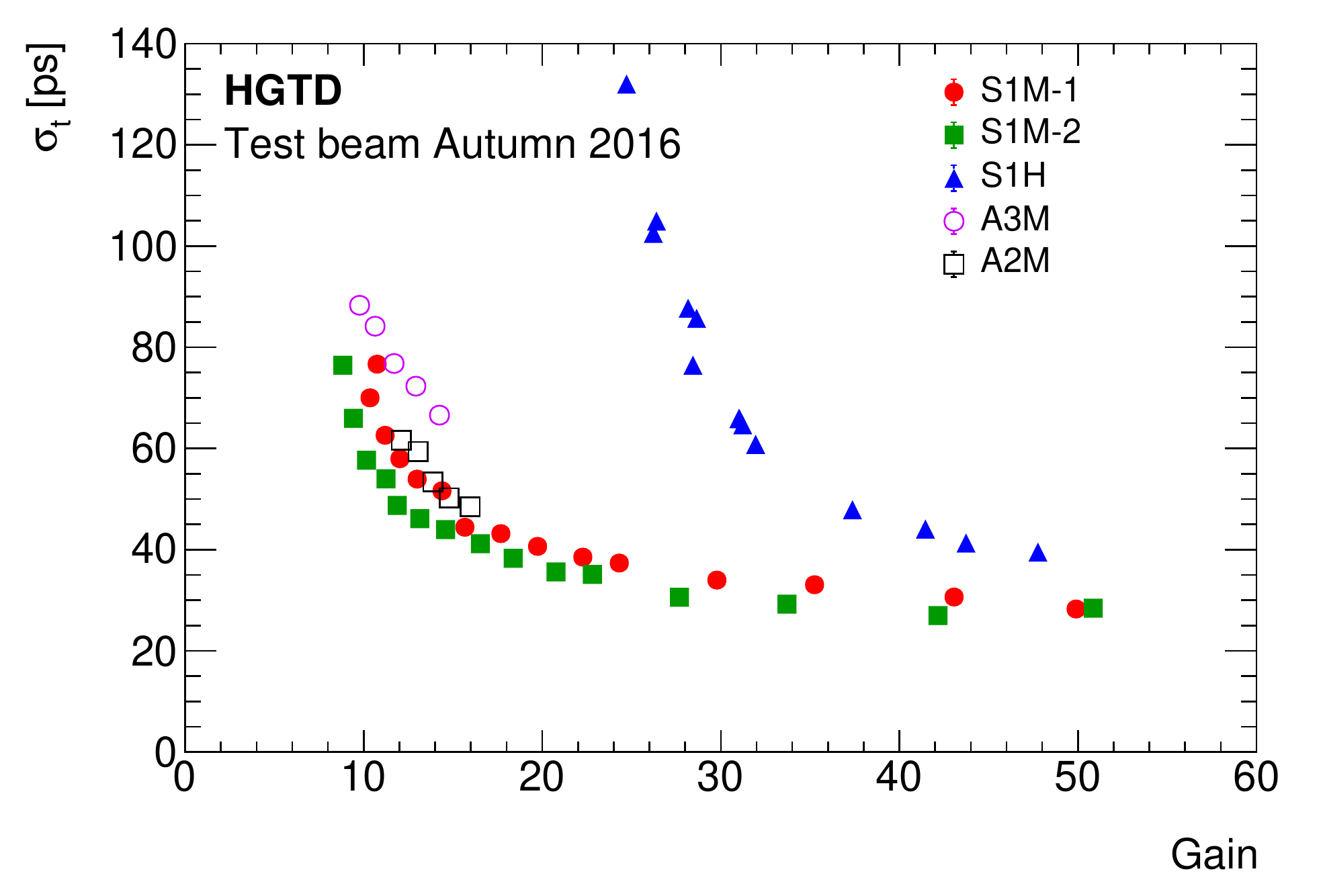}
    \label{fig:timeReso}
  }
\hfill
\subfloat[]{
    \includegraphics[width=0.45\textwidth]{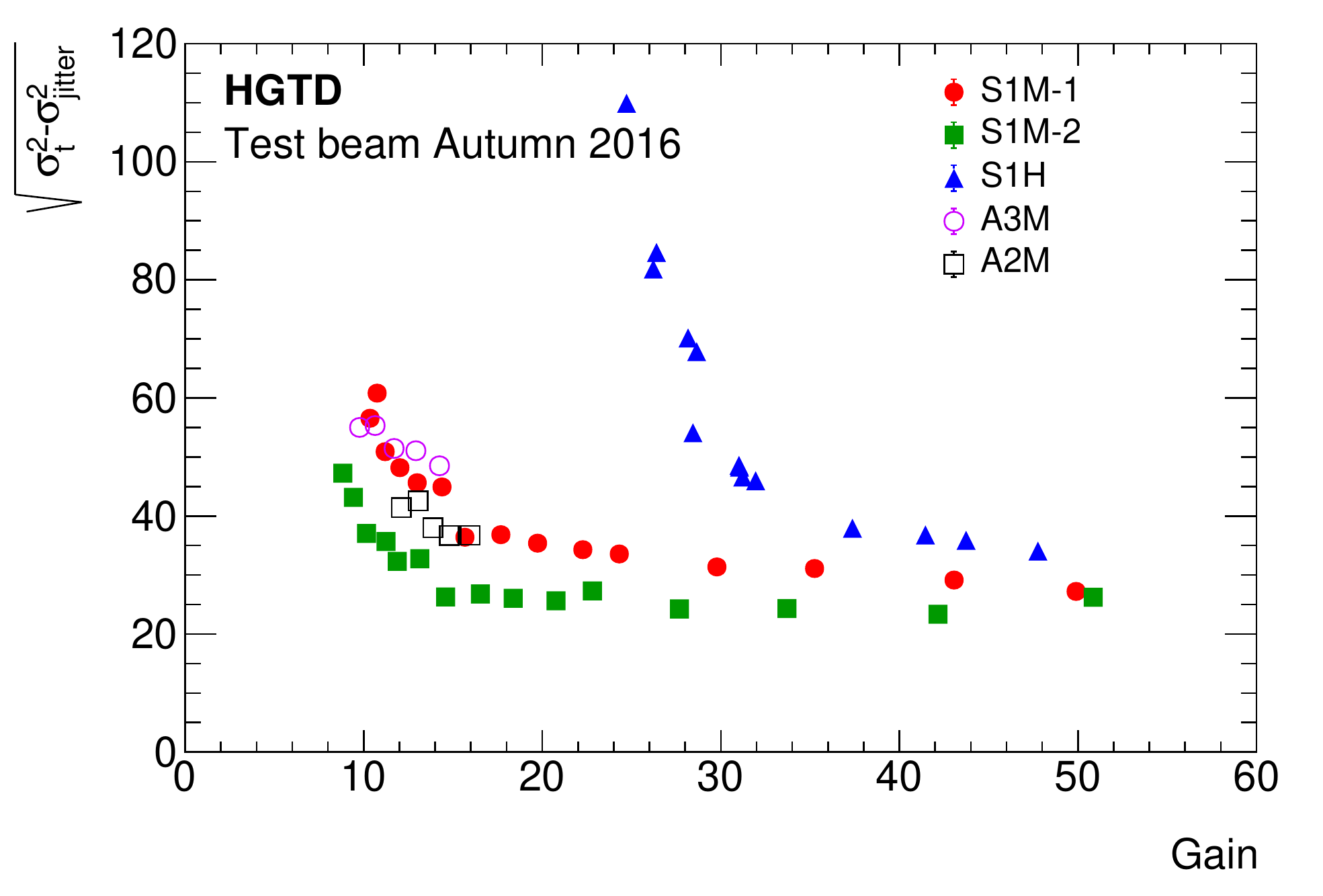}
    \label{fig:Landau}
  }

  \caption{Time resolution using the ZCD method \protect\subref{fig:timeReso} and $\sigma_{t}^{ZCD}\ominus\sigma_{jitter}^{ZCD}$ \protect\subref{fig:Landau} as a function of the gain for single pad sensors and arrays.  Statistical uncertainties are negligible and smaller than the marker size.}
  \label{fig:timeResoLandau}
\end{figure}

Figure~\ref{fig:timeReso} shows the time resolution for the ZCD method ($\sigma_{t}^{ZCD}$) as a function of the gain for single pads and arrays.
 Two approximately universal behaviours are observed for sensors with medium and high doping and, at a given gain, the best performance is obtained for sensors with medium doping.
For a gain around 14, the best time resolution is (44.0$\pm$0.5)\,ps for S1M-2, while a worse resolution is obtained for sensors with larger pad size as expected from Figure~\ref{fig:Jitter}: (53.4$\pm$0.6)\,ps for A2M and (66.6$\pm$0.4)\,ps for A3M.
In addition to the larger noise, the performance of the arrays is limited by the lowest achievable gain due to the reduced breakdown voltage. The best time resolution (27~ps) is obtained at the largest gain reached by single pad sensors.

The quadratic difference between the measured time resolution and the electronic jitter ($\sigma_{t}^{ZCD}\ominus\sigma_{jitter}^{ZCD}$) is shown in Figure~\ref{fig:Landau}. Assuming that the time resolution has only two contributions, the Landau fluctuation ($\sigma_{\mathrm{Landau}}$) and the electronic jitter, this difference should be equal to $\sigma_{\mathrm{Landau}}$.
As expected from the simulation~\cite{WF2,TrentoCartiglia}, a plateau is reached at the higher gain where the Landau fluctuations is the dominant contribution to the time resolution.
For small gain values the difference is not constant anymore: it increases, showing that additional contributions are present, whose origin has not yet been identified.


\subsubsection{Uniformity}

\begin{figure}[htbp]
  \centering
    \includegraphics[width=0.55\textwidth]{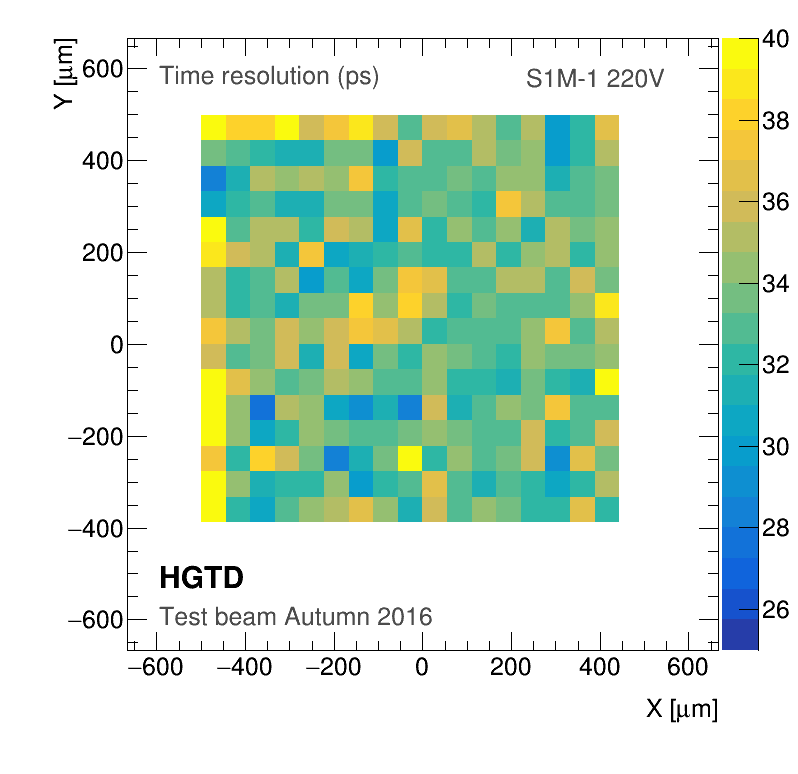}
    \label{fig:ST31_Pos}


  \caption{Time resolution in ps for single-pad sensor S1M-1 as a function of the position in the pad with a bias voltage of 220\,V.
 There is a minimum of 100 events in each bin of the size of $(55~\SI{}{\micro\meter})^2 $.
}

  \label{fig:Stime_vs_pos}
\end{figure}

For single pads, the time resolution is calculated by measuring the width of the time difference between
a  given sensor and the  fast SiPM for which the time resolution was  measured
with the test beam data to be 10.9\,ps.
The distribution of the time resolution as a function of the position on the pad is shown as a two-dimensional map in Figure~\ref{fig:Stime_vs_pos} for S1M-1 at 220 V.
The bin size is (55.5\,\SI{}{\micro\meter})$^2$ and only bins with at least 100 events are considered.
The uncertainty is below 2\,ps in each  bin and the time resolution is uniform within a few picoseconds over the DUT\footnote{For this voltage the time resolution variation as a function of the gain is small, explaining why the circular shape is not observed}.

A position dependent measurement of the time resolution was not possible for arrays due to the poor alignment of the sensors that only allowed for a small overlap region that could be traversed by the same particles.



\label{sec:TimeRes}

\FloatBarrier

\section{Conclusion}
\label{sec:conclusion}

Several measurements on single LGAD sensors with a surface of 1.3$\times$1.3\,mm$^2$ and arrays with 2$\times$2 pads with a surface of 2$\times$2\,mm$^2$ or 3$\times$3\,mm$^2$ each have been obtained from data collected during a beam test campaign in autumn 2016 with a pion beam of 120\,GeV energy at the CERN SPS. All sensors had the same thickness of 50\,\SI{}{\micro\meter}, but different implantation doses in the multiplication layer. Gain, efficiency and time resolution have been measured inclusively and as a function of the position of the particle inside the sensor by using a beam telescope with a position resolution of few micrometers. The efficiency is uniform within 1\% over the surface of each pad and the time resolution within 2\,ps. Furthermore, the fraction of events with hits above threshold in two neighbouring pads of the arrays was found to be well below 1\%.  Based on the efficiency measurement, the size of the active area was estimated and found to be compatible with the expectation from the sensor production. The sensors with a medium doping dose show better performance than the ones with a high dose and the sensors with a small surface have, as expected, a better time resolution than the larger ones. The tested non-irradiated sensors with a surface of 1.3$\times$1.3\,mm$^2$ fulfil the requirements of a time resolution of about 30\,ps for a gain of 35 (down to 27\,ps at a gain of 50) and are considered good candidates to be used for the ATLAS High Granularity Timing Detector. The time resolution was measured with different algorithms yielding consistent results. In particular, the suitable ones for implementation in the read-out electronics have similar performance as the optimal one that can only be used offline. 

\section*{Acknowledgements}

The authors gratefully acknowledge CERN and the SPS staff for successfully operating the North Experimental Area and for continuous supports to the users. We acknowledge the expert contributions of the SCIPP technical staff. Part of this work has been performed within the framework of the CERN RD50 Collaboration. 

The work was supported by: the United States Department of Energy, grant DE-FG02-04ER41286; the MINECO, Spanish Government, under grants FPA2013-48308-C2-1-P, FPA2014-55295-C3-2-R, FPA2015-69260-C3-2-R, FPA2015-69260-C3-3-R (co-financed with the European Union's FEDER funds) and SEV-2012-0234 (Severo Ochoa excellence programme), as well as under the Juan de la Cierva programme; the Spanish ICTS Network MICRONANOFABS partially supported by MINECO; the Catalan Government (AGAUR): Grups de Recerca Consolidats (SGR 2014 1177); the European Union’s Horizon 2020 Research and Innovation programme under Grant Agreement no. 654168 (AIDA-2020); the Cluster of Excellence
Precision Physics, Fundamental Interactions and Structure of Matter (PRISMA – EXC 1098) of the German Research Foundation.

\bibliographystyle{JHEP}
\bibliography{atlas-jinst}

\end{document}